\documentclass[journal,draftclsnofoot,onecolumn,12pt]{IEEEtran}
\IEEEoverridecommandlockouts
\usepackage{amsmath,amsfonts}
\usepackage{algorithm}
\usepackage{array}
\usepackage[caption=false,font=normalsize,labelfont=sf,textfont=sf]{subfig}
\usepackage{textcomp}
\usepackage{stfloats}
\usepackage{url}
\usepackage{verbatim}
\usepackage{graphicx}
\usepackage{cite}
\usepackage{booktabs}
\usepackage{threeparttable}
\usepackage{pifont}
\usepackage{algorithmicx}
\usepackage{pdfcomment}
\usepackage{bookmark}
\usepackage{enumerate}
\usepackage{bm}
\usepackage{epsfig,amssymb,color}

\newcommand{\reffig}[1]{Fig. \ref{#1}}

\usepackage{multirow}
\newcommand{\tabincell}[2]{\begin{tabular}{@{}#1@{}}#2\end{tabular}}  

\pdfoutput=1

\begin{document}

\title{Multi-Scenario Broadband Channel Measurement and Modeling for Sub-6 GHz RIS-Assisted Wireless Communication Systems}

\author{Jian Sang, Mingyong Zhou, Jifeng Lan, Boning Gao, Wankai Tang, Xiao Li,\\ Shi Jin, Ertugrul Basar, Cen Li, Qiang Cheng, and Tie Jun Cui
\thanks{Jian Sang, Mingyong Zhou, Jifeng Lan, Boning Gao, Wankai Tang, Xiao Li, and Shi Jin are with the National Mobile Communications Research Laboratory, Southeast University, Nanjing 210012, China.}
\thanks{Ertugrul Basar is with the Communications Research and Innovation Laboratory (CoreLab), Department of Electrical and Electronics Engineering, Koç University, Sariyer, Istanbul 34450, Turkey.}
\thanks{Cen Li is with the China Communications Technology Co., Ltd, Nanjing 210012, China.}
\thanks{Qiang Cheng, and Tie Jun Cui are with the State Key Laboratory of Millimeter Waves, Southeast University, Nanjing 210012, China.}
}


\maketitle

\vspace{-1cm}
\begin{abstract}
Reconfigurable intelligent surface (RIS)-empowered communication, has been considered widely as one of the revolutionary technologies for next generation networks. However, due to the novel propagation characteristics of RISs, underlying RIS channel modeling and measurement research is still in its infancy and not fully investigated. In this paper, we conduct multi-scenario broadband channel measurements and modeling for RIS-assisted communications at the sub-6 GHz band. The measurements are carried out in three scenarios covering outdoor, indoor, and outdoor-to-indoor (O2I) environments, which suffer from non-line-of-sight (NLOS) propagation inherently. Three propagation modes including intelligent reflection with RIS, specular reflection with RIS and the mode without RIS, are taken into account in each scenario. In addition, considering the cascaded characteristics of RIS-assisted channel by nature, two modified empirical models including floating-intercept (FI) and close-in (CI) are proposed, which cover distance and angle domains. The measurement results rooted in 2096 channel acquisitions verify the prediction accuracy of these proposed models. Moreover, the propagation characteristics for RIS-assisted channels, including path loss (PL) gain, PL exponent, spatial consistency, time dispersion, frequency stationarity, etc., are compared and analyzed comprehensively. These channel measurement and modeling results may lay the groundwork for future applications of RIS-assisted communication systems in practice.
\end{abstract}
 

\section{Introduction}
\label{sec1}
With the deployment of the fifth generation (5G) communication networks around the world since 2019, the global mobile data traffic has experienced a dramatic growth, which is anticipated to reach 5016 exabytes per month (Eb/mo) in 2030 \cite{ba1}. The evolutionary and advanced sixth generation (6G) communication networks with the even higher requirements, have aroused a surge of interest from both academia and industry nowadays. By 6G networks, many emerging application services such as virtual reality, holographic telepresence, pervasive connectivity, etc., are envisioned promisingly to be realized \cite{ba2, baa2}. Up to now, although no detailed specifications have been laid out on the communication standard of 6G, the study of possible innovating technologies, applicable propagation model and practical channel measurement facing 6G, have been the focus of initiatives in various countries \cite{ba3, baa3}. 

Reconfigurable intelligent surface (RIS)-empowered communication, with its ability of controlling the propagation environment effectively, has recently been considered one of the revolutionary technologies to meet the ever-increasing 6G demands \cite{ba6, ba8}. RIS provides a prospective concept by manipulating the electromagnetic (EM) wave which is impinging on it and reflected to the desired direction, with a relatively low power consumption. By this means, RIS can constructively customize wireless channels instead of traditionally adapting to the harsh propagation environment \cite{v1_b2}. 


Generally speaking, an RIS is composed of an artificial meta-surface with plentiful sub-wavelength sized unit cells arranged in an array form. The reflection coefficients of each unit cell, e.g., magnitude, phase, polarization, etc., can be reconfigured intelligently through adjusting the control voltage \cite{ba10}. Typically, each unit cell is comprised of conductive printed patch, which has a size of proportion to the wavelength of operating frequency. In order to reconfigure the EM response, a tunable load, such as positive-intrinsic-negative (PIN) diode or varactor diode, could be embedded into unit cell to interact with control voltage. In this way, the EM characteristics of the outgoing signal can be programmed independently by a smart controller such as field programmable gate array (FPGA) in real time. Thanks to such innovative characteristics brought by RISs, the radio propagation environment may be redefined \cite{ba11, ba12}.

Until now, numerous analyses on RIS-empowered systems have been theoretically investigated in many fields, including but not limited to: user localization \cite{ba13}, secure communication \cite{ba14}, beam optimization \cite{bv4}, user scheduling \cite{v1_b1}, etc. Meanwhile, theoretical channel modeling as well as practical measurements, on power gain and coverage enhancement of RIS-assisted non-line-of-sight (NLOS) communications have gradually been reported \cite{bv5, bs1}. Nevertheless, the real channel measurement campaign and applicable channel model in practice, which describe the underlying physical propagation phenomena of EM waves incurred by RIS, are still in their infancy and rarely proposed, despite their critical roles. 


\subsection{Modeling-related works}
Considering broadcast and beamforming scenarios respectively, \cite{b4} and \cite{b5} theoretically formulated the path loss (PL) expressions for RIS-assisted NLOS communications in free space. Then a two-path propagation model for RIS-assisted communications was proposed in \cite{b6}, which showed that the deployment of RIS was advantageous for mitigating fast fading. In \cite{bs2}, a physics-driven PL model for RIS-assisted link in free space was proposed based on antenna theory. In \cite{bs3}, the PL scaling law of the scattered field was discussed numerically, which unified the opposite behavior of RIS serving as a scatterer and a mirror. \cite{bs4} formulated the RIS-related PL models by using the vector generalization of Green’s theorem. In \cite{bs5}, a PL model for RIS-assisted systems was proposed based on physical optics technique, which explained the mechanism why the plentiful unit cells on RIS individually acted as diffuse scatterers but could jointly perform beamforming in a desired direction. \cite{bs6} considered that RIS were partitioned into several tiles, which yielded the free-space tile response model for RIS-assisted link. In \cite{bv1}, an RIS channel simulator was introduced by considering the 3GPP channel modeling methodology for millimeter wave (mmWave) channels. 

\subsection{Measurement-related works}
In \cite{b4}, Tang {\em et al.} conducted the measurements to validate PL models in an anechoic chamber, utilizing three RISs at 4.25 GHz and 10.5 GHz respectively. Further mmWave measurements on two customized RIS covering 27 GHz and 33 GHz were carried out in \cite{b5} to verify the refined PL models. In \cite{bss7}, two RISs constituted of varactor diodes and PIN diodes respectively, were employed in an indoor environment to illustrate the RIS-related channel reciprocity. By utilizing universal software radio peripherals (USRPs), the measurement campaigns in \cite{bs9} illustrated that an RIS could bring a power gain of 27 dB and 14 dB for short-range and long-range communications, respectively. In \cite{bs10}, a signal-to-noise ratio (SNR) gain of 8 dB was shown to be offered by the RIS at 5.8 GHz when the NLOS distance reached 35 m. \cite{bs14} proposed a spectrum sharing solution using RIS in indoor channel, where experiments conducted at 2.4 GHz demonstrated a higher spectrum-spatial efficiency by controlling the phase shifts of RIS. A 1-bit RIS fabricated in \cite{b10} validated its capability of enhancing power gains by conducting the indoor measurements at 3.5 GHz. In \cite{b11}, the experiments conducted on a multi-bit RIS substantiated a power improvement of 2.65 dB compared with 1-bit RIS. In \cite{bs12}, the RIS experiments at 28 GHz manifested its capability for transmission signal enhancement in NLOS links. Considering the RIS serving as transmitter (Tx), the measurements at 28 GHz in \cite{bs11} illustrated that the PL in the main beam directions were improved by RIS. \cite{bv2} considered practical measurements at 5 GHz and employed RIS codebooks to dynamically adjust the RIS phases depending on user locations. In terms of physical layer security, measurements were performed in \cite{bv3}, which showed that with RIS deployment, the signal level of an eveasdropper could be reduced down to the noise floor.



Though the above-mentioned studies have largely developed RIS-based free-space models and power-related measurements, unabridged channel observations in practical environment are still rarely reported. Meanwhile, accurate, realistic, yet easy-to-use RIS-empowered channel models are eagar to be developed. Undoubtedly, the RIS-assisted cascaded channel exhibits many new propagation characteristics, such as multi-dimensional PL, diversified propagation modes, unpredictable spatial evolution, etc., which greatly differ from the traditional propagation channels. In consequence, unlike only considering the power gain or simplistically verifying the free-space PL model in existing works, generalizable channel modeling and real channel measurement in practical RIS-related communication scenarios are desired to be fully investigated.

\subsection{Main contributions}

Motivated by such circumstances, in this paper, we conduct multi-scenario broadband channel measurements and channel modeling for RIS-assisted single-input single-output (SISO) communications. Considering the dominating role of the sub-6 GHz band serving as the 5G commercial operating frequency band \cite{bs19}, the assessment of RIS-assisted channel at this band will probably be of great significance for its future applications, because of the interoperability between RIS and 5G networks. Thus, we carry out the RIS-assisted channel measurements at the sub-6 GHz licensed frequency band. From multiple perspectives covering frequency domain, spatial domain, and temporal domain, we investigate the propagation characteristics of RIS-assisted channels and explain their underlying phenomena. The major contributions are listed as follows.


\begin{itemize}

\item We conduct broadband channel measurement campaigns at 2.6 GHz for RIS-assisted NLOS propagation in three scenarios, covering outdoor, indoor, and outdoor-to-indoor (O2I) measurements. Three propagation modes including intelligent reflection with RIS, specular reflection with RIS, and the mode without RIS, are considered in each scenario. In addition, the measured channel data is also compared with the free-space propagation model. The multi-dimensional parameters including Tx-RIS distance, receiver (Rx)-RIS distance, angle-of-arrival (AoA) onto RIS, and angle-of-departure (AoD) from RIS, etc., are taken into account in our measurement.  

\item We propose the modified empirical floating-intercept (FI) and close-in (CI) models, to cater for the multi-parameter characteristics of RIS-assisted channels. In comparison to the CI and FI models that we previously proposed in \cite{bs15}, we further refine them from single-dimensional variable to multi-dimensional variables, with distance information as well as angle information included. 

\item The measurement results are presented and the modeling accuracy is analyzed. From multiple perspectives covering PL gain, PL exponent (PLE), spatial consistency, time dispersion, frequency stationarity, etc., we investigate the characteristics of RIS-assisted channels under different modes as well as in different scenarios. Their underlying propagation phenomena such as the claim of ``customizing wireless channels'', are explained and several potential research directions are also pointed out.  

\end{itemize}

\subsection{Organization}

The remainder of this paper is outlined as follows. In Section II, the measurement systems and the deployed RIS are described in detail. The measurement scenarios and the measurement procedures are provided in Section III. Then the post-processing of measurement data and the proposed channel models are illustrated in Section IV. In Section V, the measurement results and modeling analyses are presented. Finally, conclusions are drawn and several future works are laid out in Section VI.

\vspace{-0.2cm}
\section{Channel Measurement Systems and the Deployed RIS}
\label{sec2}

\begin{figure}[htbp]
\vspace{-0.5cm}
\centering
\begin{minipage}[t]{0.48\textwidth}
\centering
\includegraphics[width=2.6in]{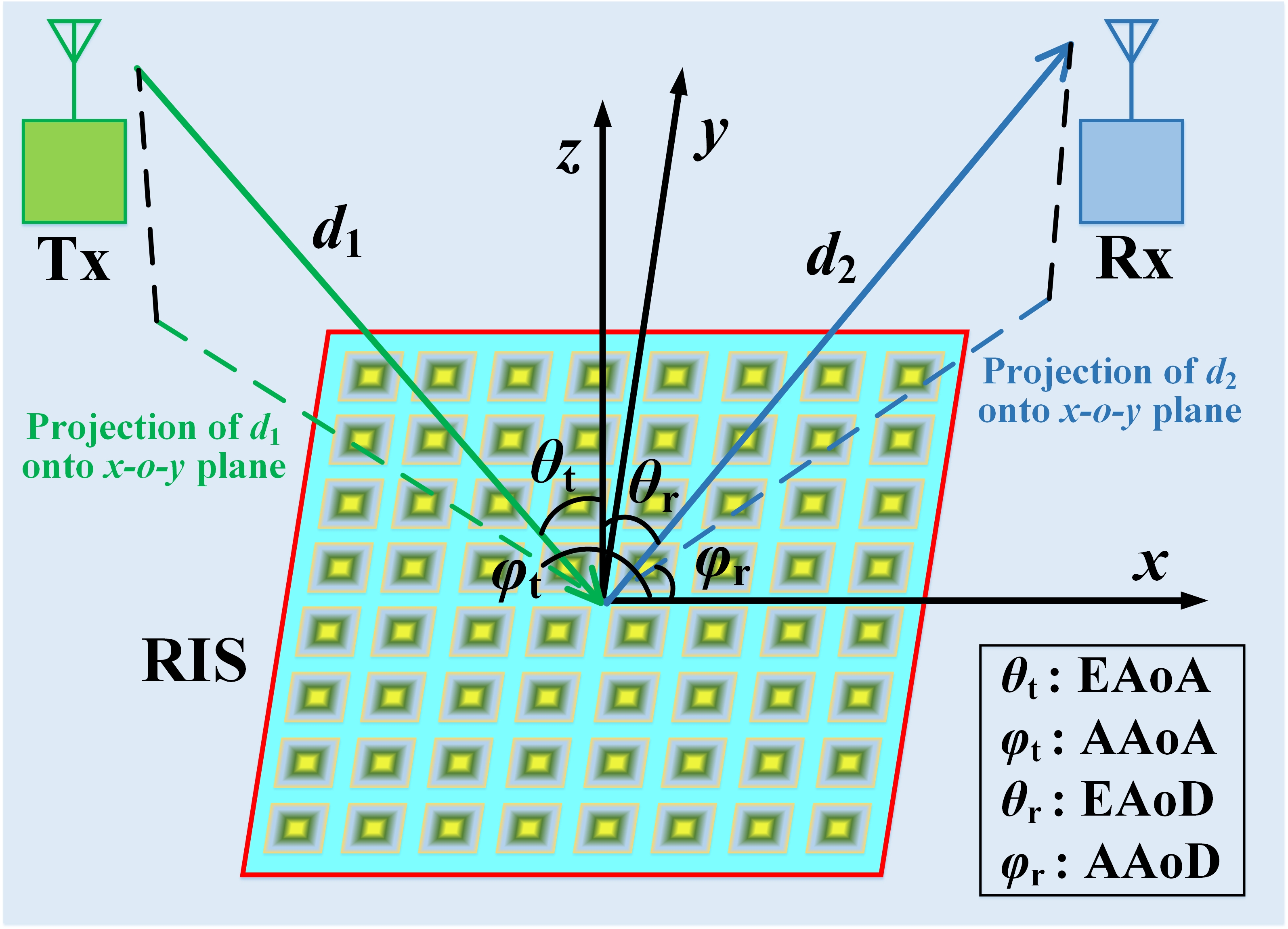}
\vspace{-0.5cm}
\caption{Parameter definitions for RIS-assisted channel.}
\label{fig1}
\end{minipage}
\begin{minipage}[t]{0.48\textwidth}
\centering
\includegraphics[width=2.5in]{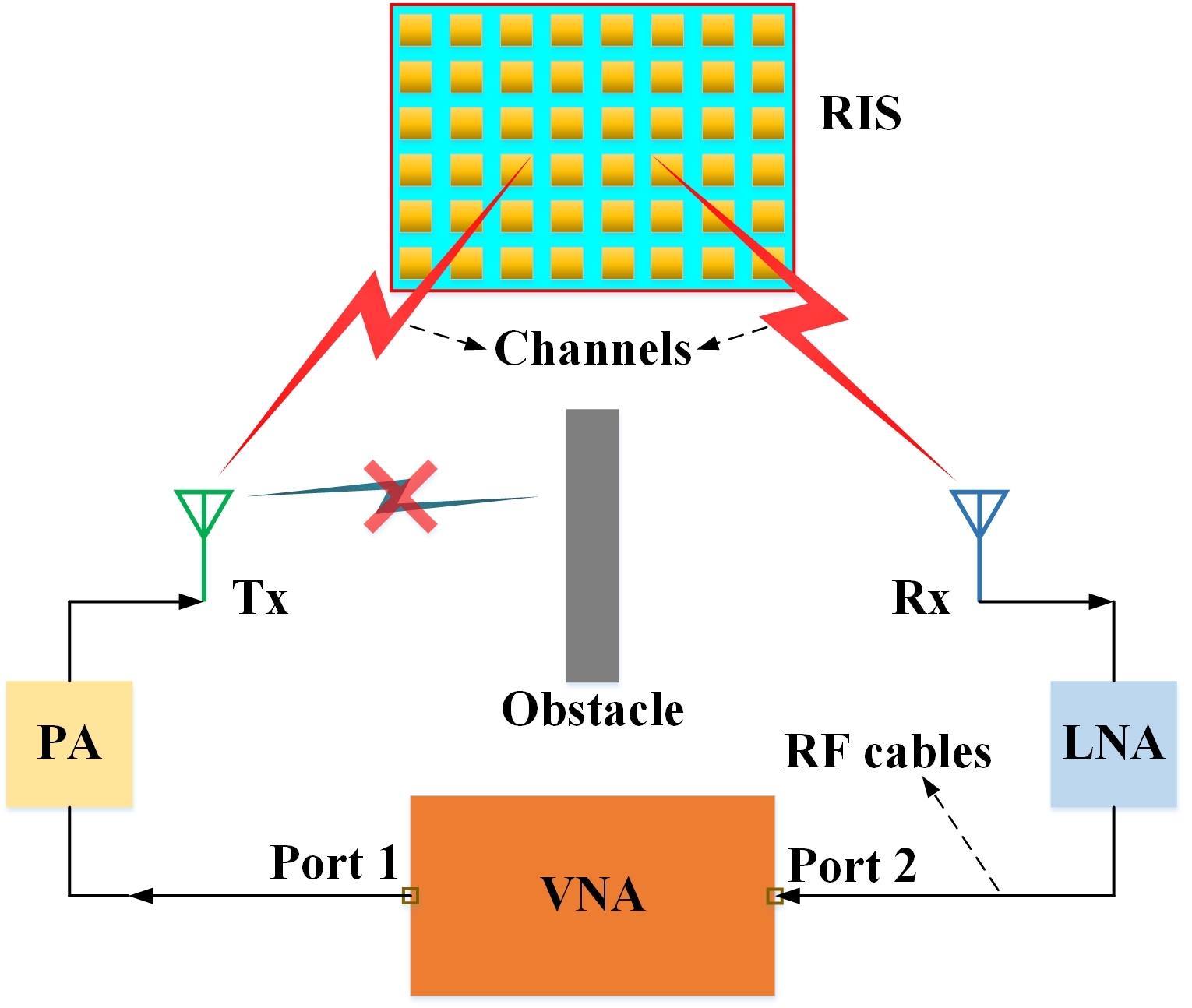}
\vspace{-0.5cm}
\caption{Descriptions of channel measurement systems.}
\label{fig2}
\end{minipage}
\vspace{-1cm}
\end{figure}

In this paper, an RIS-assisted SISO communication system is considered, as shown in \reffig{fig1}. The direct channel between Tx and Rx is blocked by obstacles, i.e., a NLOS channel. An RIS is deployed to assist the signal propagation, which is referred to the RIS-assisted channel with cascaded LOS link in the rest of this paper. The RIS consists of $M$ unit cells per row and $N$ unit cells per column, i.e., $M \times N$ unit cells in total. Each unit cell has a rectangular shape with a horizontal size of $d_x$ and a vertical size of $d_y$. Moreover, each unit cell has a constant reflection magnitude $A$ and programmable binary discrete phases. Note that the RIS in this paper refers to the family of reflective RIS, by which the signal is neither amplified nor retransmitted. As shown in \reffig{fig1}, let $d_1$ and $d_2$ refer to the distances from Tx to the center of RIS and from Rx to the center of RIS, respectively. $\theta_t$, $\varphi_t$, $\theta_r$ and $\varphi_r$ respectively indicate the elevation angle-of-arrival (EAoA) and the azimuth angle-of-arrival (AAoA) onto RIS as well as the elevation angle-of-departure (EAoD) and the azimuth angle-of-departure (AAoD) from RIS.

\subsection{Channel measurement systems}

As shown in \reffig{fig2}, the channel measurement systems are composed of a VNA, a fabricated RIS, a power amplifier (PA), a low noise amplifier (LNA), radio frequency (RF) cables, and two directional horn antennas. In detail, the used VNA version is Agilent N5230C PNA-L. The bandwidth of $190$ MHz ranging from $2.5$ GHz to $2.69$ GHz is selected for the measured broadband signal, which covers the licensed frequency band of current commercial mobile networks \cite{bs19}. There are $191$ scanning points within this measurement bandwidth, with scanning time of $27.3$ $\mu$s for each point by the VNA. As shown in \reffig{fig2}, the port 1 of VNA is used to transmit signal and its port 2 receives the signal. Therefore, the scanning data ${S_{21}}$ is collected as channel transfer function (CTF) in frequency domain. In addition, the antennas used in the measurement system are two wideband double-ridged horn antennas, which serve as Tx and Rx respectively. Each of them is vertically polarized with a gain of $8.25$ dBi and has a half power beam width (HPBW) of $60$ degrees. The PA connected to Tx has a $30$ dB gain and the LNA connected to Rx has a $24$ dB gain. The low-loss RF cables with stable amplitude and phase responses are utilized to connect the above equipment. The detailed configurations and parameters of the measurement systems are summarized in \autoref{table1}.

\begin{table}[htbp]
\vspace{-0.5cm}
\caption{Configurations and parameters of measurement systems.}
\label{table1}
    \centering
     \vspace{-0.2cm}
\begin{tabular}{|c|c|c|c|}
\hline
\textbf{Configuration}  & \textbf{Parameter} & \textbf{Configuration}  & \textbf{Parameter}  \\ \hline
Measurement bandwidth & $190$ MHz, $2.5\sim2.69$ GHz &  Transmitted power of VNA        & $10$ dBm  \\ \hline
Number of frequency scanning points & $191$ & Antenna gain of Tx/Rx  & $8.25$ dBi \\ \hline
Scanning time for one measurement     & $0.0273$ ms$\times$ $191$ $\approx$ $5.21$ ms  &  Antenna HPBW of Tx/Rx  & $60^\circ$    \\ \hline
Antenna polarization of Tx/Rx  & vertical  & Loss of RF cables  & $\le$ $0.5$ dB/m \\\hline                
PA gain & $30$ dBm  & LNA gain  & $24$ dBm  \\ \hline
Size of RIS  & $1.6$ m $\times$ $0.8$ m & Number of unit cell  & $32\times16 = 512$     \\ \hline
Size of unit cell  & $0.05$ m $\times$ $0.05$ m  & Phase resolution of unit cell        & 1-bit        \\ \hline
\multirow{2}{*}{\tabincell{c}{Phase levels of unit cell}} & \multirow{2}{*}{\tabincell{c}{\{ $55^\circ$ for coding ``0'',\vspace{-0.2cm}\\ $-125^\circ$ for coding ``1''\}}} & \multirow{2}{*}{\tabincell{c}{Phase difference of unit cell}} & \multirow{2}{*}{\tabincell{c}{$180^\circ$}}  \\ 
& & & \\ \hline
\end{tabular}
\vspace{-0.8cm}
\end{table}

\vspace{-0.2cm}
\subsection{The deployed RIS}
\label{sec2.2}

\begin{figure*}[htbp]
\vspace{-0.5cm}
\centering
\includegraphics[width=5.5in]{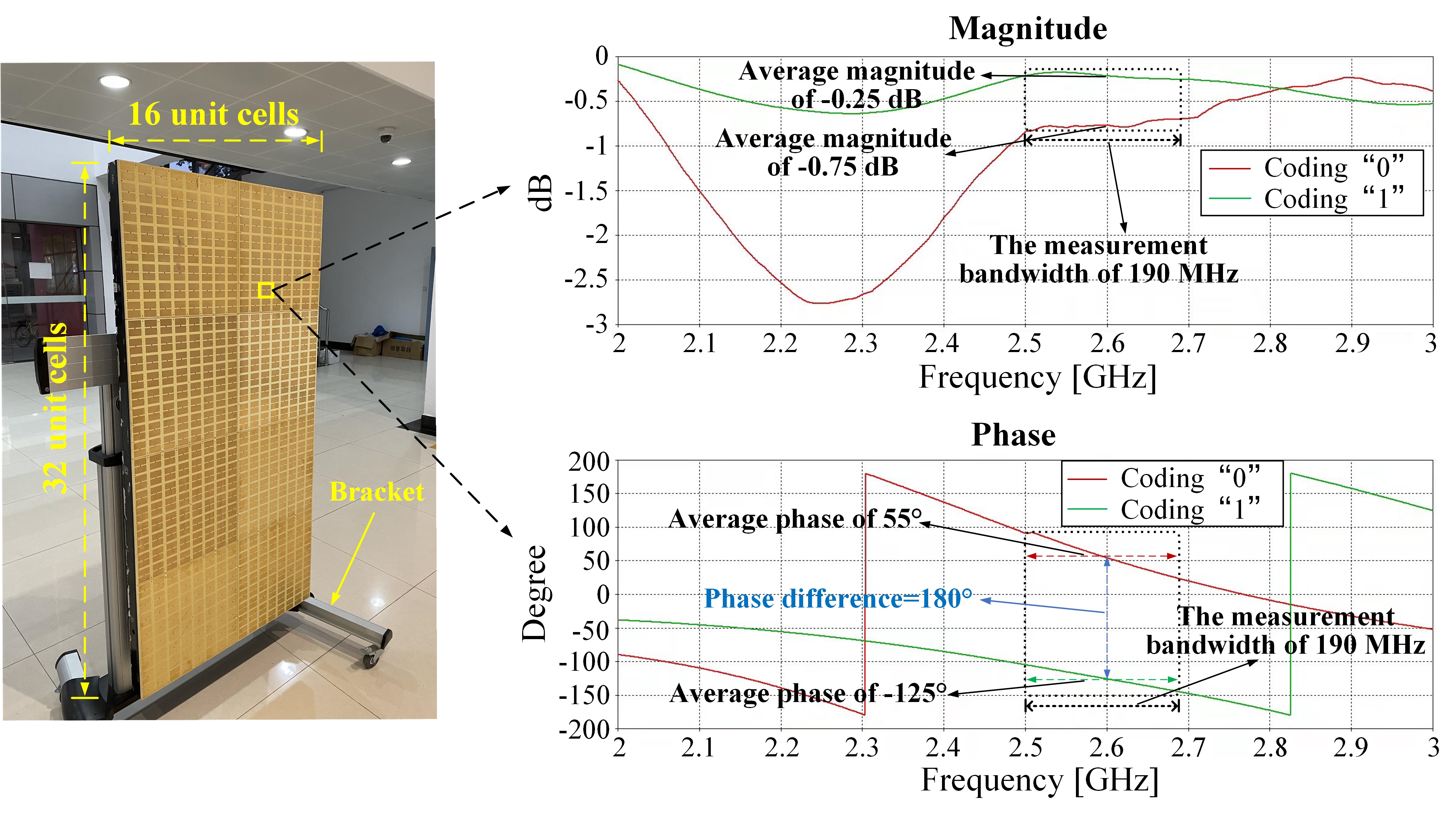}
\vspace{-0.5cm}
\caption{Image and response characteristics of the deployed RIS.}
\label{fig3}
\vspace{-0.8cm}
\end{figure*}

As displayed in \reffig{fig3}, a fabricated RIS with the central operating frequency of $2.6$ GHz is utilized for channel measurements. This RIS is composed of $32$ unit cells per column and $16$ unit cells per row, i.e., $512$ unit cells in total, which has a physical size of $1.6$ m $\times$ $0.8$ m. Each unit cell on the RIS has a square shape of $0.05$ m $\times$ $0.05$ m, which is nearly half-wavelength of the central operating frequency. A PIN diode is embedded into the unit cell to tune the response characteristics of impinging EM waves. In addition, each unit cell can be independently programmed with 1-bit phase resolution, i.e., two coding states of coding ``0'' and coding ``1'', which correspond to the PIN diode being “off” state and “on” state, respectively. 

\reffig{fig3} also shows the magnitude and phase response curves of this RIS with respect to the operating frequency. Within the measurement bandwidth from $2.5$ GHz to $2.69$ GHz, the magnitude has an average loss of $0.25$ dB for coding ``1'' and an average loss of $0.75$ dB for coding ``0'', respectively. It is worth mentioning that these losses are much lower than the gain to be provided by RIS beamforming. In addition, within the measurement bandwidth, the phase has an average value of $55^\circ$ for coding ``1'' and an average value of $-125^\circ$ for coding ``0''. Thus, a phase difference of $180^\circ$ between these two coding states can be achieved.

By pre-designing the coding of each unit cell, this RIS can perform arbitrary beamforming towards desired azimuth and elevation directions. Note that in this paper, the coding schemes of RIS are designed by \textbf{``Dynamic Threshold Phase Quantization (DTPQ) method''} in \cite{bs17}, which can be regarded as the optimal scheme in free space and applicable to various communication scenarios. In our measurement campaigns, firstly the parameters $d_1$, $d_2$, $\theta_t$, $\varphi_t$, $\theta_r$ and $\varphi_r$ of the RIS-assisted channels at each point are measured and calculated respectively, according to the geometric position. Based on these parameters, we utilize the ``DTPQ method'' to generate the codebooks of RIS, which are mapped to each measured point one-to-one. Then, the channel data at each measured point is collected with RIS configuring the corresponding codebook.

\vspace{-0.2cm}
\section{Channel Measurement Scenarios and Procedures}
\label{sec3}

\begin{figure*}[htbp]
\vspace{-0.8cm}
\centering
     \subfloat[]{\includegraphics[height = 1.3in]{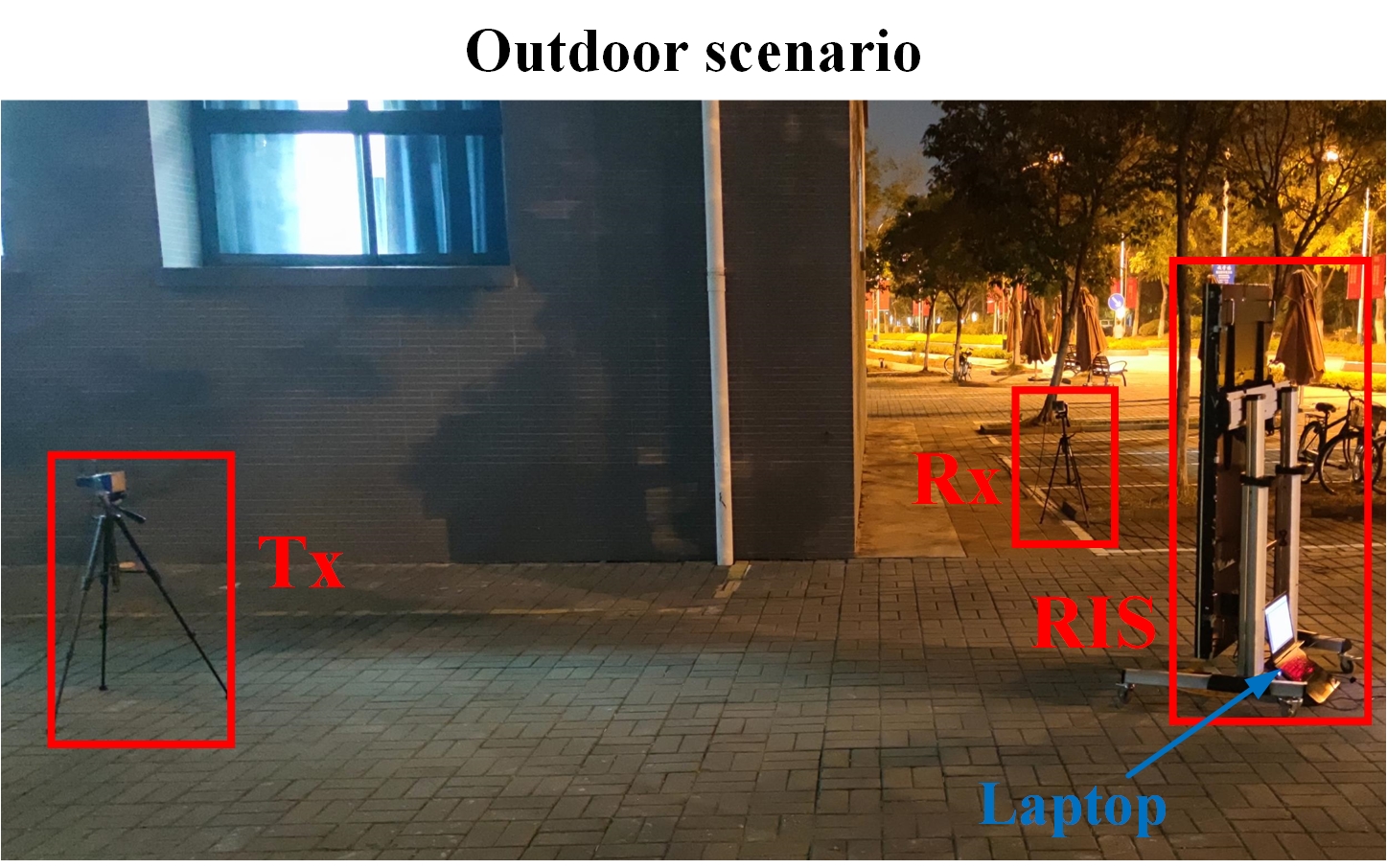}\label{scenario_a}}\hfill
     \subfloat[]{\includegraphics[height = 1.3in]{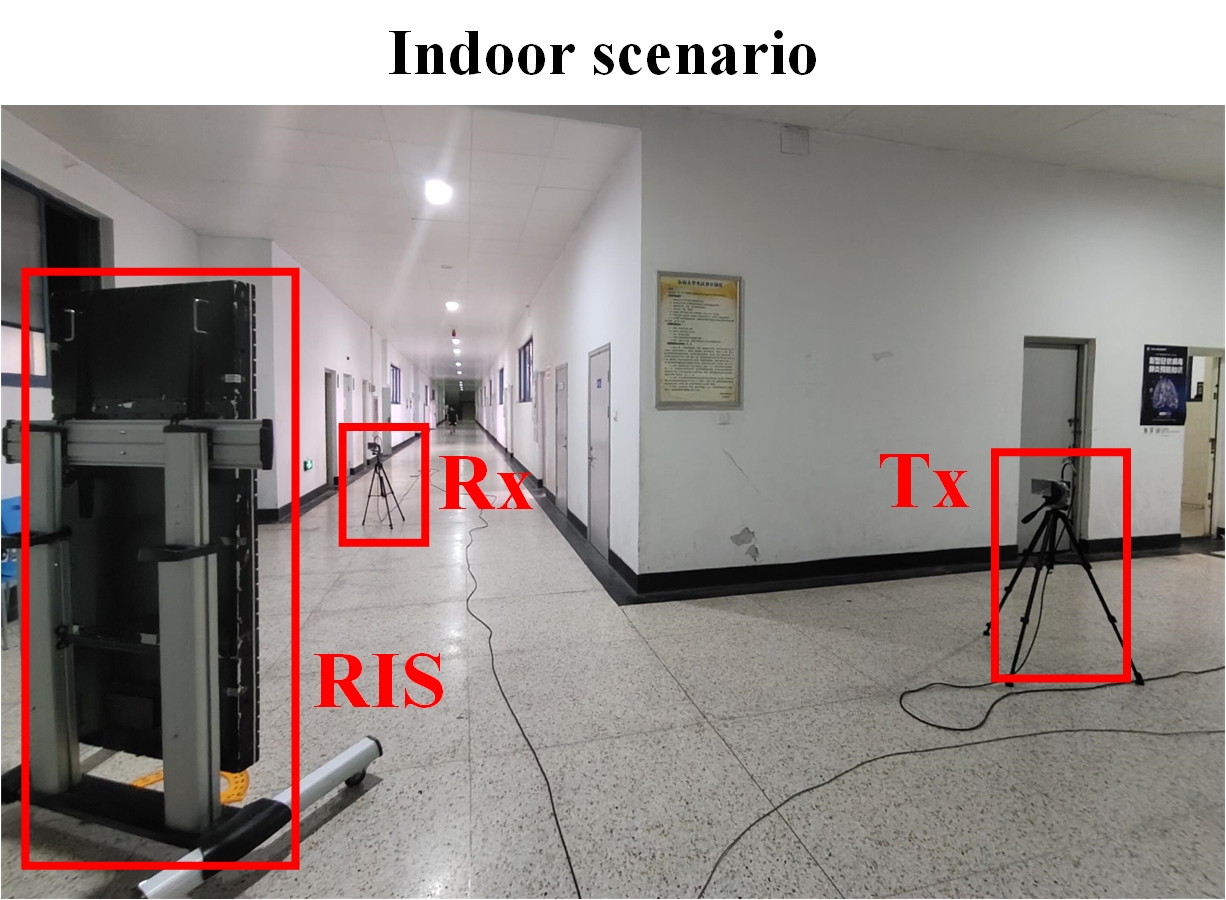}\label{scenario_b}}\hfill
     \subfloat[]{\includegraphics[height = 1.3in]{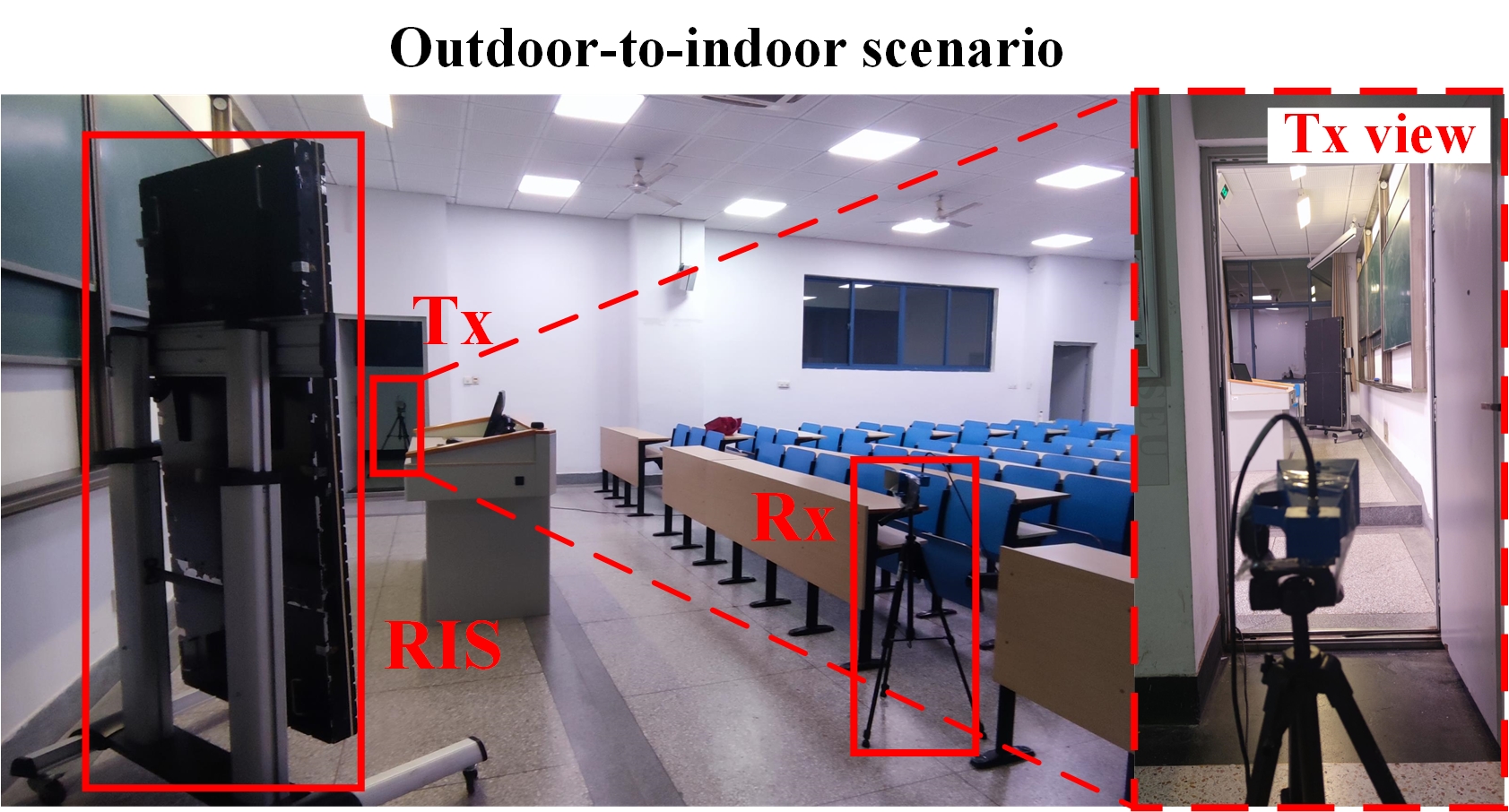}\label{scenario_c}}\hfill
     \vspace{-0.2cm}
    \caption{Images of three channel measurement campaigns. (a) Outdoor scenario. (b) Indoor scenario. (c) O2I scenario.}
\label{fig4}
\vspace{-0.2cm}
\end{figure*}

In order to comprehensively investigate the propagation characteristics of RIS-assisted channel in practical communications, three typical scenarios including outdoor, indoor, and O2I environments are selected to conduct channel measurement, which cover $2096$ channel acquisitions to be measured in total. These measurement scenarios are located at Jiulonghu Campus of Southeast University, in Nanjing, China. The images of channel measurement campaigns in these scenarios are respectively exhibited in \reffig{fig4}. Our measurement campaigns are conducted at night so as to avoid the influences of pedestrians or vehicles on the measurement results. Thus, the measured channels can be viewed as quasi-static. In particular, the detailed information on measurement scenarios and procedures are provided in the subsequent subsections.

\begin{figure}[htbp]
\vspace{-0.2cm}
\centering
\includegraphics[width=6.4in]{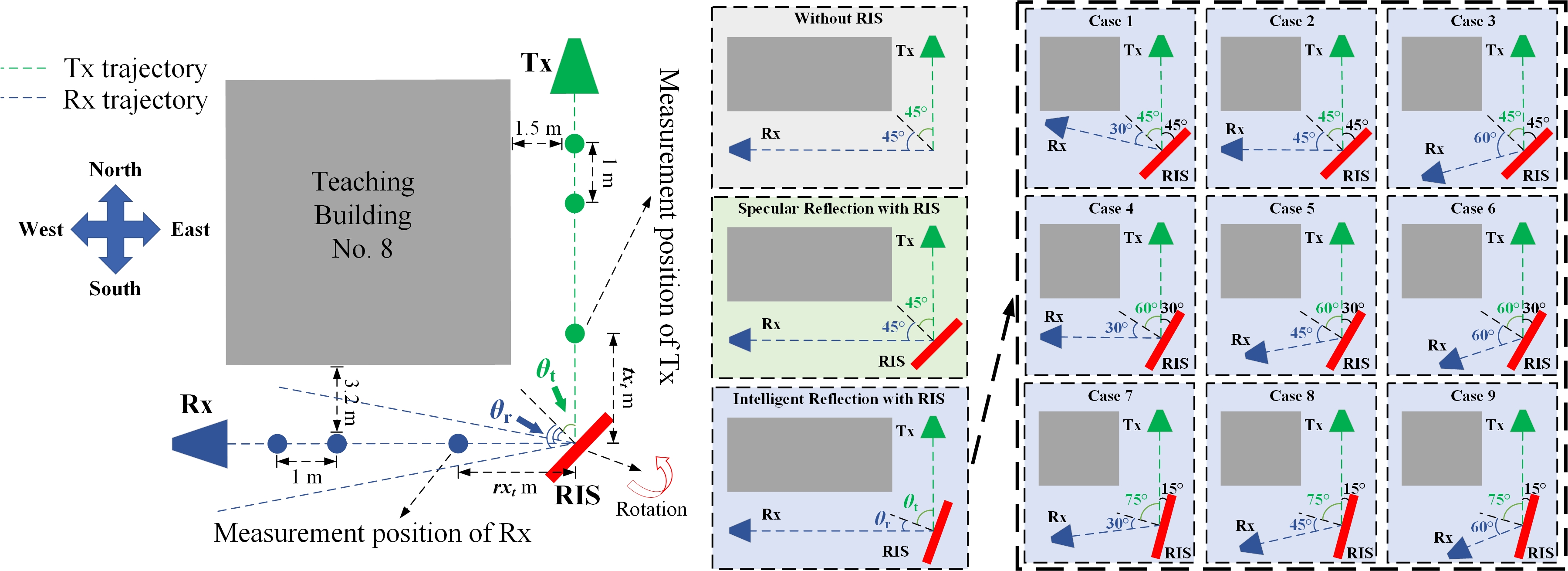}
\vspace{-0.5cm}
\caption{Scenario and procedure for outdoor measurement.}
\label{fig5}
\vspace{-0.5cm}
\end{figure}

\vspace{-0.2cm}
\subsection{Outdoor channel measurement}

For outdoor measurement, the building corner next to Teaching Building No. 8 outdoors is selected as measurement site. The scenario and procedure for this measurement are exhibited in \reffig{fig5}. As shown in this figure, the communication links between two sides of this building suffer from NLOS propagation, due to the blockage by the exterior wall of building. The exterior facades of this building are fully covered by concretes. Two horn antennas serving as Tx and Rx, are respectively placed at the east-side and south-side spaces of the building. 

From \reffig{fig5}, Tx is kept at a perpendicular distance of $1.5$ m from the east facades of building, and moves along the line (referred to as ``Tx trajectory line'') parallel to the east facades. In addition, Rx is kept at a perpendicular distance of $3.2$ m from the south facades of building, and moves along the line (referred to as ``Rx trajectory line'') parallel to the south facades. The RIS is positioned at the intersection of Tx and Rx trajectory lines. Throughout this measurement, both Tx and Rx maintain facing the RIS.

Tx is initially placed at a distance of $tx_t$ m from RIS and moves away from RIS at a step of $1$ m along the Tx trajectory line, until the Tx-RIS distance reaches $tx_p$ m. Similarly, Rx is initially placed at a distance of $rx_t$ m from the RIS and moves away from RIS at a step of $1$ m along the Rx trajectory line, until the Rx-RIS distance reaches $rx_p$ m. Note that in this measurement, Tx, Rx, and the center of RIS are fixed to be at the same height of $1$ m above the ground, thus the AAoA $\varphi_t = 180^\circ$ and the AAoD $\varphi_r = 0^\circ$ invariably.

\begin{table*}[htbp]
\vspace{-0.5cm}
\centering
\footnotesize
\caption{Measurement parameters of outdoor scenario.}
\label{table2}
 \vspace{-0.2cm}
\begin{tabular}{|c|c|c|c|c|c|c|c|c|c|c|c|}
\hline
\multirow{3}{*}{\tabincell{c}{\textbf{Propagation}\vspace{-0.2cm}\\\textbf{mode}}} 
& \multirow{3}{*}{\tabincell{c}{\textbf{Without}\vspace{-0.2cm}\\\textbf{RIS}}} 
& \multirow{3}{*}{\tabincell{c}{\textbf{Specular}\vspace{-0.2cm}\\\textbf{reflection} \vspace{-0.2cm}\\\textbf{with RIS}}} & \multicolumn{9}{|c|}{\multirow{1}{*}{\tabincell{c}{\textbf{Intelligent reflection with RIS}}}} \\ 
\cline{4-12}
& & & \multirow{2}{*}{\tabincell{c}{Case\vspace{-0.2cm}\\1}}& \multirow{2}{*}{\tabincell{c}{Case\vspace{-0.2cm}\\2}} & \multirow{2}{*}{\tabincell{c}{Case\vspace{-0.2cm}\\3}} & \multirow{2}{*}{\tabincell{c}{Case\vspace{-0.2cm}\\4}} & \multirow{2}{*}{\tabincell{c}{Case\vspace{-0.2cm}\\5}} & \multirow{2}{*}{\tabincell{c}{Case\vspace{-0.2cm}\\6}} & \multirow{2}{*}{\tabincell{c}{Case\vspace{-0.2cm}\\7}} & \multirow{2}{*}{\tabincell{c}{Case\vspace{-0.2cm}\\8}} & \multirow{2}{*}{\tabincell{c}{Case\vspace{-0.2cm}\\9}} \\ 
 & & & & & & & & & & & \\ \hline

EAoA $\theta_t$ & $45^\circ$ \textit{(virtual)} & $45^\circ$ & $45^\circ$ & $45^\circ$ & $45^\circ$ & $60^\circ$ & $60^\circ$ & $60^\circ$ & $75^\circ$ & $75^\circ$ & $75^\circ$ \\ \hline
EAoD $\theta_r$ & $45^\circ$ \textit{(virtual)} & $45^\circ$ & $30^\circ$ & $45^\circ$ & $60^\circ$ & $30^\circ$ & $45^\circ$ & $60^\circ$ & $30^\circ$ & $45^\circ$ & $60^\circ$ \\ \hline
$tx_t$ (m) & $5$ & $5$ & $9$ & $5$ & $9$ & $9$ & $9$ & $9$ & $9$ & $9$ & $9$  \\ \hline
$tx_p$ (m) & $18$ & $18$ & $18$ & $18$ & $18$ & $18$ & $18$ & $18$ & $18$ & $18$ & $18$  \\ \hline
$rx_t$ (m) & $5$ & $5$ & $5$ & $5$ & $8$ & $8$ & $8$ & $8$ & $8$ & $8$ & $8$  \\ \hline
$rx_p$ (m) & $18$ & $18$ & 12 & $18$ & $15$ & $18$ & $15$ & $15$ & $15$ & $15$ & $15$  \\ \hline

   \multirow{2}{*}{\tabincell{c}{Number of \vspace{-0.2cm}\\ Tx points}} 
& \multirow{2}{*}{\tabincell{c}{$14$}} 
& \multirow{2}{*}{\tabincell{c}{$14$}} 
& \multirow{2}{*}{\tabincell{c}{$10$}} 
& \multirow{2}{*}{\tabincell{c}{$14$}} 
& \multirow{2}{*}{\tabincell{c}{$10$}} 
& \multirow{2}{*}{\tabincell{c}{$10$}} 
& \multirow{2}{*}{\tabincell{c}{$10$}} 
& \multirow{2}{*}{\tabincell{c}{$10$}} 
& \multirow{2}{*}{\tabincell{c}{$10$}} 
& \multirow{2}{*}{\tabincell{c}{$10$}} 
& \multirow{2}{*}{\tabincell{c}{$10$}} \\
 & & & & & & & & & & & \\ \hline

  \multirow{2}{*}{\tabincell{c}{Number of \vspace{-0.2cm}\\ Rx points}} 
& \multirow{2}{*}{\tabincell{c}{$14$}} 
& \multirow{2}{*}{\tabincell{c}{$14$}} 
& \multirow{2}{*}{\tabincell{c}{$8$}} 
& \multirow{2}{*}{\tabincell{c}{$14$}} 
& \multirow{2}{*}{\tabincell{c}{$8$}} 
& \multirow{2}{*}{\tabincell{c}{$11$}} 
& \multirow{2}{*}{\tabincell{c}{$8$}}
& \multirow{2}{*}{\tabincell{c}{$8$}}
& \multirow{2}{*}{\tabincell{c}{$8$}}
& \multirow{2}{*}{\tabincell{c}{$8$}}
& \multirow{2}{*}{\tabincell{c}{$8$}} \\
 & & & & & & & & & & &   \\ \hline

\multirow{2}{*}{\tabincell{c}{Number of \vspace{-0.2cm}\\ total points}} 
& \multirow{2}{*}{\tabincell{c}{$14\times14$\vspace{-0.2cm}\\$= 196$}} 
& \multirow{2}{*}{\tabincell{c}{$14\times14$\vspace{-0.2cm}\\$= 196$}}  
&\multicolumn{9}{|c|}{\multirow{2}{*}{\tabincell{c}{ $10\times8+14\times14+10\times8+10\times11+10\times8+10\times8+$\vspace{-0.2cm}\\ $10\times8+10\times8+10\times8 = 866$}}}    \\ 
& & & \multicolumn{9}{|c|}{} \\ \hline
\end{tabular}
\vspace{-0.5cm}
\end{table*}

In this measurement, the channels under three propagation modes are measured, including \textit{specular reflection with RIS}, \textit{intelligent reflection with RIS}, and \textit{without RIS}. These three modes include $196$, $866$, and $196$ points to be measured respectively. The detailed configurations on these three modes are parameterized in \autoref{table2}, which are illuminated as follows.

For the mode of specular reflection with RIS, the RIS is fixed at $\theta_t = 45^\circ$ and $\theta_r = 45^\circ$. $tx_t$, $tx_p$, $rx_t$, and $rx_p$ are $5$ m, $18$ m, $5$ m, and $18$ m, respectively. In this mode, the RIS does not perform phase regulation, and all unit cells of RIS are uniformly configured to be coding ``0''. Thus, the RIS can be regarded to be equivalent to an equal-sized metal plate \cite{b5}.

For the mode of intelligent reflection with RIS, the RIS performs phase regulation according to the coding scheme mentioned in Section \ref{sec2.2}. In this mode, RIS is appropriately rotated to form $9$ different pairs of EAoA $\theta_t$ and EAoD $\theta_r$, i.e., ${\theta _t} \in \{45^\circ, 60^\circ, 75^\circ\}$ and ${\theta _r} \in \{30^\circ, 45^\circ, 60^\circ\}$, as illustrated in \reffig{fig5}. 

For the mode without RIS, the RIS is removed. Thus, there are a virtual EAoA $\theta_t = 45^\circ$ and a virtual EAoD $\theta_r = 45^\circ$. $tx_t$, $tx_p$, $rx_t$, and $rx_p$ are $5$ m, $18$ m, $5$ m, and $18$ m, respectively.

\begin{figure}[htbp]
\vspace{-0.5cm}
\centering
\includegraphics[width=6in]{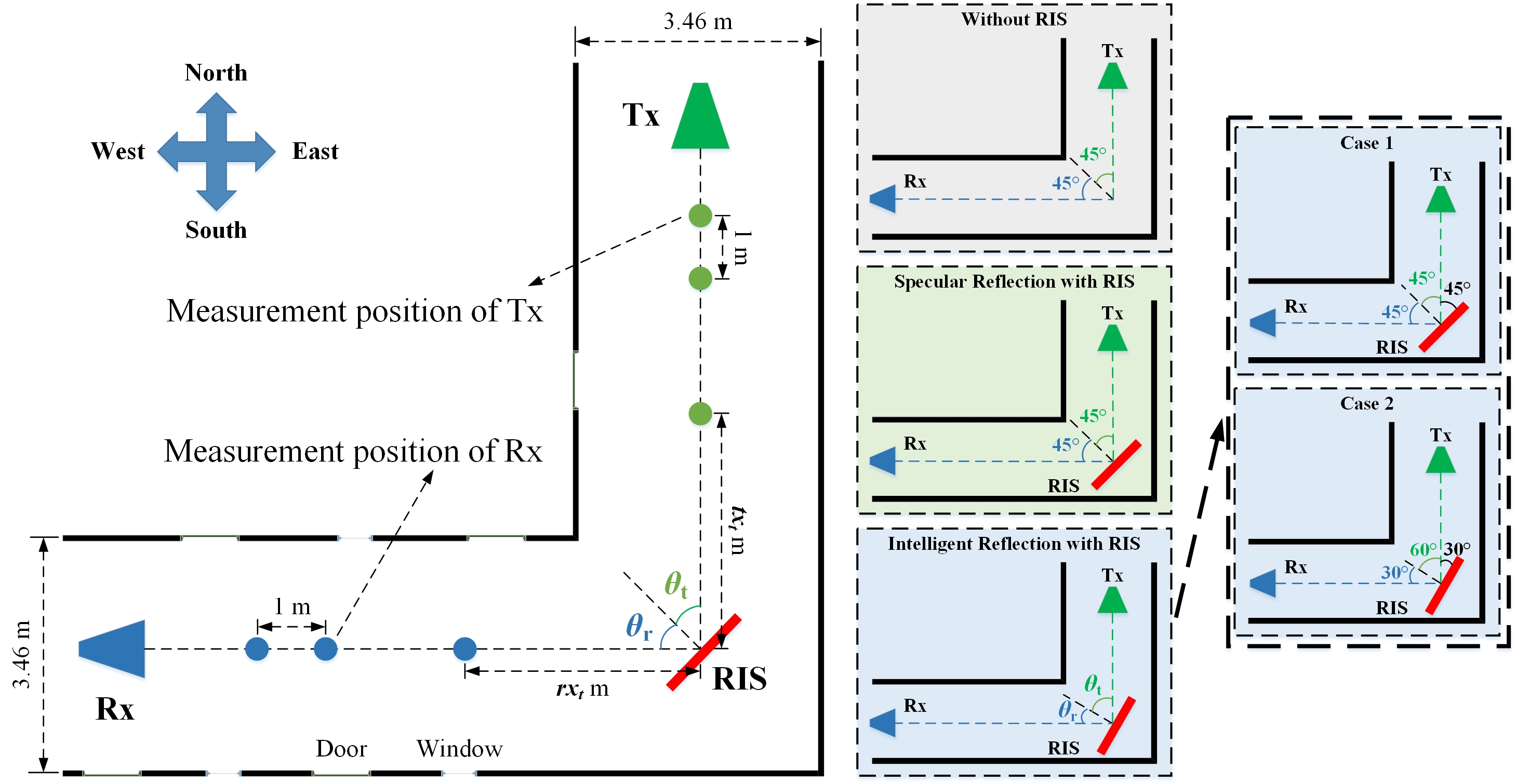}
\vspace{-0.5cm}
\caption{Scenario and procedure for indoor measurement.}
\label{fig6}
\vspace{-0.5cm}
\end{figure}

\subsection{Indoor channel measurement}
For indoor measurement, the corridor on the first floor of Teaching Building No. 2, which is next to the Liangjiang Road, is selected as measurement site. The measurement scenario and procedure are shown in \reffig{fig6}. From this figure, an inherent NLOS transmission link exists, due to the fact that both sides of the corridor are perpendicular to each other. The walls of this corridor are covered with concrete, among which the aluminum alloy doors and several glazed windows are embedded. The corridor has a width of $3.46$ m and a height of $3.32$ m. The south-north corridor and the east-west corridor are respectively selected to place Tx and Rx, and the RIS is employed at the corner of corridor. Two horn antennas serving as Tx and Rx are respectively placed facing the RIS throughout this measurement.

As shown in \reffig{fig6}, Tx moves along the central axis of south-north corridor (referred to as ``Tx trajectory line'') and Rx moves along the central axis of east-west corridor (referred to as ``Rx trajectory line''). The RIS is positioned at the intersection of Tx and Rx trajectory lines. Tx is initially placed at a distance of $tx_t$ m from RIS and moves away from RIS at a step of $1$ m along the Tx trajectory line, until the Tx-RIS distance reaches $tx_p$ m. Similar with the outdoor scenario, Rx is initially placed at a distance of $rx_t$ m from the RIS and moves away from RIS at a step of $1$ m along the Rx trajectory line, until the Rx-RIS distance reaches $rx_p$ m. In this measurement, Tx, Rx, and the center of RIS are also fixed to be at the same height of $1$ m above the ground, thus the AAoA $\theta_t = 180^\circ$ and the AAoD $\theta_r = 0^\circ$ invariably.

In this measurement, three propagation modes as in the outdoor scenario, including \textit{specular reflection with RIS}, \textit{intelligent reflection with RIS}, and \textit{without RIS}, are also measured. These three modes contain $196$, $392$, and $196$ points to be measured respectively. The detailed configurations of the three modes are summarized in \autoref{table3}, which are further illuminated as follows.

\begin{table}[htbp]
\vspace{-0.5cm}
\centering
\footnotesize
\caption{Measurement parameters of indoor scenario.}
\label{table3}
 \vspace{-0.2cm}
\begin{tabular}{|c|c|c|c|c|}
\hline
\multirow{2}{*}{\tabincell{c}{\textbf{Propagation mode}}} 
& \multirow{2}{*}{\tabincell{c}{\textbf{Without RIS}}} 
& \multirow{2}{*}{\tabincell{c}{\textbf{Specular reflection with RIS}}} & \multicolumn{2}{|c|}{{\textbf{Intelligent reflection with RIS}}}  
\\ \cline{4-5}
& & & Case 1 & Case 2  \\ \hline

EAoA $\theta_t$ & $45^\circ$ \textit{(virtual)} & $45^\circ$ & $45^\circ$ & $60^\circ$ \\ \hline
EAoD $\theta_r$ & $45^\circ$ \textit{(virtual)} & $45^\circ$ & $45^\circ$ & $30^\circ$ \\ \hline
$tx_t$ (m) & $5$ & $5$ & $5$ & $5$     \\ \hline
$tx_p$ (m) & $18$ & $18$ & $18$ & $18$ \\ \hline
$rx_t$ (m) & $5$ & $5$ & $5$ & $5$     \\ \hline
$rx_p$ (m) & $18$ & $18$ & $18$ & $18$ \\ \hline
Number of Tx points & $14$ & $14$ & $14$ & $14$ \\ \hline
Number of Rx points & $14$ & $14$ & $14$ & $14$ \\ \hline
Number of total points & $14\times14=196$ & $14\times14=196$
& \multicolumn{2}{|c|}{$14\times14+14\times14=392$} \\ \hline
\end{tabular}
\vspace{-0.5cm}
\end{table}

For the mode of specular reflection with RIS, the RIS is fixed at $\theta_t = 45^\circ$ and $\theta_r = 45^\circ$. Meanwhile, $tx_t$, $tx_p$, $rx_t$, and $rx_p$ are $5$ m, $18$ m, $5$ m, and $18$ m, respectively. In this mode, the RIS does not perform phase regulation and all unit cells of RIS are uniformly configured to be coding ``0''. For the mode of intelligent reflection with RIS, the RIS performs phase regulation according to the coding scheme mentioned in Section \ref{sec2.2}. In this mode, RIS is appropriately rotated to form 2 different pairs of $\theta_t$ and $\theta_r$, i.e., $\theta_t = 45^\circ$, $\theta_r = 45^\circ$ and $\theta_t = 60^\circ$, $\theta_r = 30^\circ$, as illustrated in \reffig{fig6}. For the mode without RIS, measurement settings are the same with the corresponding mode in the outdoor scenario, as shown in \autoref{table3}.

\vspace{-0.2cm}
\subsection{O2I channel measurement}

\begin{figure}[htbp]
\vspace{-0.5cm}
\centering
\includegraphics[width=5.5in]{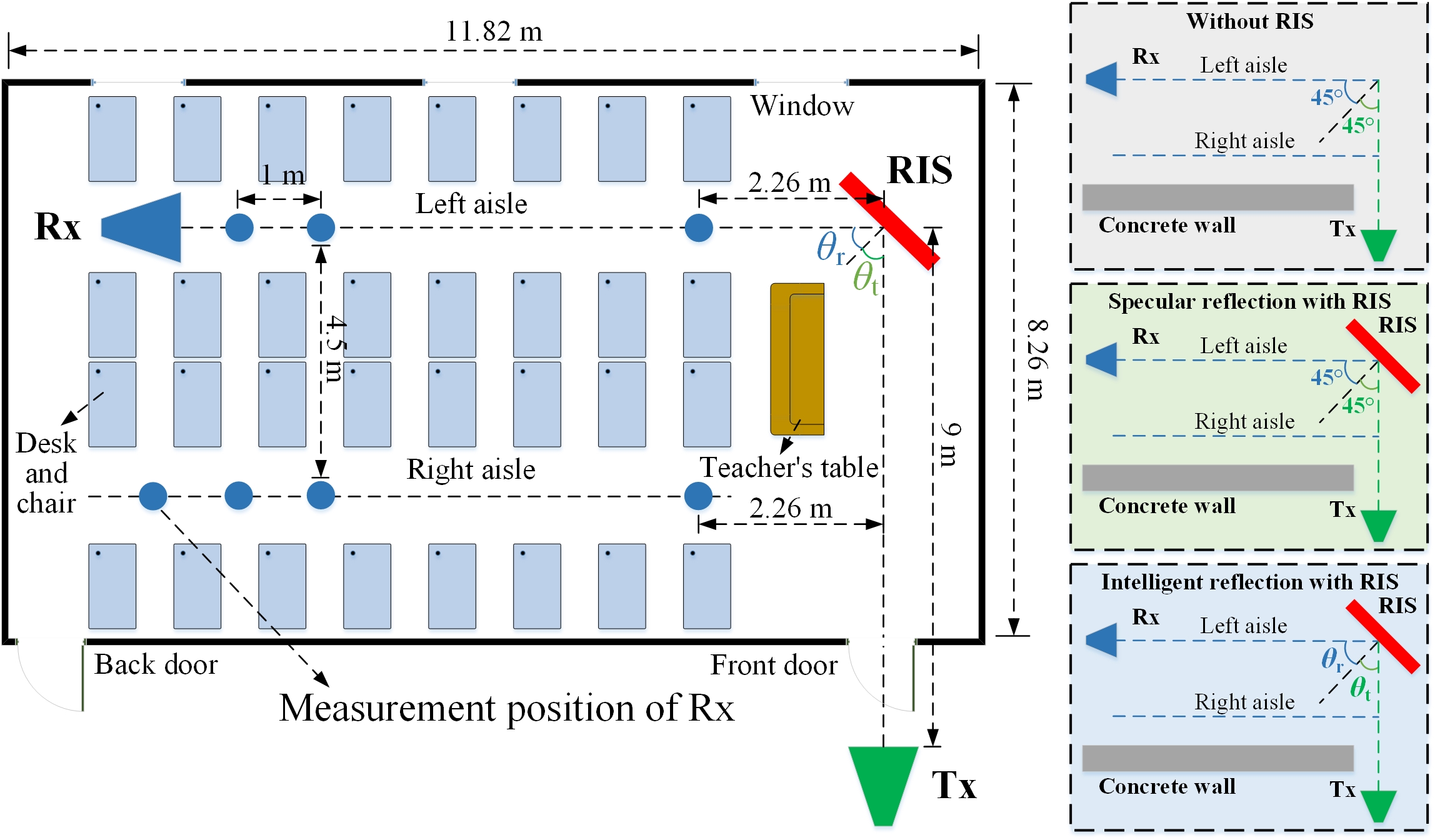}
\vspace{-0.5cm}
\caption{Scenario and procedure for O2I measurement.}
\label{fig7}
\vspace{-0.5cm}
\end{figure}

For O2I measurement, the classroom on the first floor of Teaching Building No. 7, is selected as measurement site. The scenario and procedure for this measurement are exhibited in \reffig{fig7}. In this scenario, Rx in the classroom is weakly covered by communication signals when Tx locates outside the classroom, due to the blockage by concrete walls between Tx and Rx. Thus, their communication links can be reasonably considered as NLOS. This measured classroom has a width of $8.26$ m, a length of $11.82$ m, and a height of $3.32$ m. It is surrounded by concrete walls, where several glazed windows and two aluminum alloy doors are embedded into the walls. In addition, this classroom is typically filled with a number of orderly-arranged desks and chairs, as well as a teacher’s table. There are two aisles in the classroom, as shown in \reffig{fig7}. 

In this measurement, Tx is fixed outside the classroom and faces the front door, which is kept open all along. The RIS is positioned at the intersection of the Tx-front door line and the left aisle, with a distance of $9$ m away from Tx and $\theta_t = 45^\circ$. Two horn antennas serving as Tx and Rx maintain facing the RIS throughout this measurement. Rx moves along these two aisles, respectively. In left aisle, $\theta_r = 45^\circ$ can be readily determined at each measured point. Rx is initially placed at a distance of $rx_t = 2.26$ m from RIS and moves away from RIS at a step of $1$ m, until the final distance reaches $rx_p = 10.26$ m. In right aisle, Rx is initially positioned at a perpendicular distance of $rx_t = 2.26$ m to the Tx-RIS line, and moves away from this line at a step of $1$ m, until the final distance reaches $rx_p = 10.26$ m. Based on geometric position relationships, the EAoD $\theta _r^i$ at the $i$th point in right aisle can be expressed as 
\begin{equation}
\label{eq2}
\theta _r^i = \arctan \left( {\frac{{2.26 + (i - 1) \times 1}}{{4.5}}} \right) - \left( {\frac{\pi }{2} - {\theta _t}} \right).
\end{equation}
Then we obtain the EAoD set $\mathcal{R} = \{-18.33^\circ, -9.08^\circ, -1.57^\circ, 4.45^\circ, 9.29^\circ, 13.2^\circ, 16.42^\circ, 19.08^\circ,$ $21.32^\circ\}$ in the order of positions 1$\sim$9 in right aisle. Similar with the outdoor and indoor scenarios, in this measurement, Tx, Rx, and the center of RIS are also fixed to be at the same height of $1$ m above the ground, thus the AAoA $\varphi_t = 180^\circ$ and the AAoD $\varphi_t = 0^\circ$ invariably.

\begin{table*}[htbp]
\vspace{-0.5cm}
\centering
\footnotesize
\caption{Measurement parameters of O2I scenario.}
\label{table4}
 \vspace{-0.2cm}
\begin{tabular}{|c|c|c|c|c|c|c|}
\hline
\multirow{2}{*}{\tabincell{c}{\textbf{Propagation}\\ \textbf{mode}}} 
& \multicolumn{2}{|c|}{\multirow{1}{*}{\tabincell{c}{\textbf{Without RIS}}}}
& \multicolumn{2}{|c|}{\multirow{1}{*}{\tabincell{c}{\textbf{Specular reflection with RIS}}}}
& \multicolumn{2}{|c|}{\multirow{1}{*}{\tabincell{c}{\textbf{Intelligent reflection with RIS}}}} \\
\cline{2-7} 

& Left aisle & Right aisle & Left aisle & Right aisle & Left aisle & Right aisle \\ \hline
EAoA $\theta_t$ & $45^\circ$ \textit{(virtual)} & $45^\circ$ \textit{(virtual)} & $45^\circ$ & $45^\circ$ & $45^\circ$ & $45^\circ$ \\ \hline

EAoD $\theta_r$ & $45^\circ$ \textit{(virtual)}
& $\mathcal{R}$ \textit{(virtual)} & $45^\circ$ & $\mathcal{R}$ 
& $45^\circ$
& $\mathcal{R}$ \\ \hline
$d_1$ (m) & $9$ & $9$ & $9$ & $9$ & $9$ & $9$   \\ \hline

\multirow{3}{*}{\tabincell{c}{$rx_t$ (m)}} 
& \multirow{3}{*}{\tabincell{c}{$2.26$}}
& \multirow{3}{*}{\tabincell{c}{$2.26$ (perpendi- \vspace{-0.2cm}\\cular distance \vspace{-0.2cm}\\ to Tx-RIS line)}}
& \multirow{3}{*}{\tabincell{c}{$2.26$}}
& \multirow{3}{*}{\tabincell{c}{$2.26$ (perpendi-\vspace{-0.2cm}\\cular distance\vspace{-0.2cm}\\ to Tx-RIS line)}}
& \multirow{3}{*}{\tabincell{c}{$2.26$}}
& \multirow{3}{*}{\tabincell{c}{$2.26$ (perpendi-\vspace{-0.2cm}\\cular distance\vspace{-0.2cm}\\ to Tx-RIS line)}}\\ & & & & & & \\ & & & & & & \\ \hline

\multirow{3}{*}{\tabincell{c}{$rx_p$ (m)}} 
& \multirow{3}{*}{\tabincell{c}{$10.26$}}
& \multirow{3}{*}{\tabincell{c}{$10.26$ (perpendi-\vspace{-0.2cm}\\cular distance\vspace{-0.2cm}\\ to Tx-RIS line)}}
& \multirow{3}{*}{\tabincell{c}{$10.26$}}
& \multirow{3}{*}{\tabincell{c}{$10.26$ (perpendi-\vspace{-0.2cm}\\cular distance\vspace{-0.2cm}\\ to Tx-RIS line)}}
& \multirow{3}{*}{\tabincell{c}{$10.26$}}
& \multirow{3}{*}{\tabincell{c}{$10.26$ (perpendi-\vspace{-0.2cm}\\cular distance\vspace{-0.2cm}\\ to Tx-RIS line)}}\\ & & & & & & \\ & & & & & & \\ \hline

\multirow{2}{*}{\tabincell{c}{Number of \vspace{-0.2cm}\\Rx points}} 
& \multirow{2}{*}{\tabincell{c}{$9$}}
& \multirow{2}{*}{\tabincell{c}{$9$}}
& \multirow{2}{*}{\tabincell{c}{$9$}}
& \multirow{2}{*}{\tabincell{c}{$9$}}
& \multirow{2}{*}{\tabincell{c}{$9$}}
& \multirow{2}{*}{\tabincell{c}{$9$}}
\\ & & & & & & \\ \hline

\multirow{2}{*}{\tabincell{c}{Number of \vspace{-0.2cm}\\ total points}} 
& \multicolumn{2}{|c|}{\multirow{2}{*}{\tabincell{c}{$9+9=18$}}}
& \multicolumn{2}{|c|}{\multirow{2}{*}{\tabincell{c}{$9+9=18$}}}
& \multicolumn{2}{|c|}{\multirow{2}{*}{\tabincell{c}{$9+9=18$}}} \\
& \multicolumn{2}{|c|}{} & \multicolumn{2}{|c|}{} & \multicolumn{2}{|c|}{}  \\ \hline
\end{tabular}
\vspace{-0.5cm}
\end{table*}

In this measurement, the previously mentioned three propagation modes, i.e., \textit{specular reflection with RIS}, \textit{intelligent reflection with RIS}, and \textit{without RIS}, are measured. The detailed configurations on these three modes are summarized in \autoref{table4}. The coding schemes of RIS under each mode are the same as the corresponding ones in the outdoor and indoor scenarios. For each mode in this measurement, $18$ points are measured.


\vspace{-0.2cm}
\section{Data Post-processing and Channel Modeling}
\label{sec4}

In this section, the post-processing of measurement data including PL, PDP, root mean square delay spread (RMS DS), etc., are introduced. In addition, based on the traditionally-used FI and CI models, two empirical channel models for RIS-assisted channel covering distance information and angle information are proposed.

\vspace{-0.2cm}
\subsection{Measurement data post-processing}

For the sake of illustration, the signals transmitted from port 1 of VNA and received at port 2 of VNA in frequency domain are denoted as $X(f)$ and $Y(f)$ respectively. The measured system response and channel response are represented as $G(f)$ and $H(f)$, respectively. Denote the collected data by VNA as $H_{V}(f)$, which can be calculated by
\begin{equation}
\label{eq3}
H_{V}(f) = {S_{21}}(f) = \frac{{Y(f)}}{{X(f)}} = G(f)H(f).
\end{equation}
Through the back-to-back calibration, the system response $G(f)$ can be obtained and removed from ${S_{21}}(f)$. Then the channel response in frequency domain can be calculated by
\begin{equation}
\label{eq5}
H(f) = \frac{{H_{V}(f)}}{{G(f)}}.
\end{equation}
The PL of the measured channel in logarithmic scale can be calculated by
\begin{equation}
\label{eq6}
PL = 10 \times {\rm{log_{10}}}\left( {\frac{{\sum\nolimits_{k = 1}^K {({{\left| {{H_k}(f)} \right|}^2})} }}{K}} \right) - {G_t} - {G_r},
\end{equation}
where ${{H_k}(f)}$ denotes the channel response of the $k$th frequency scanning point, $K=191$ is the number of frequency scanning points, $G_t$ and $G_r$ denote the antenna gains of Tx and Rx in logarithmic scale, respectively. 

Subsequently, the channel impulse response (CIR) in temporal domain can be determined by inverse Fourier transform,
\begin{equation}
\label{eq7}
h(t,\tau ) = {\rm{IFT}}\left( {H(f) \times {{\rm{W}}_{{\rm{hann}}}}} \right),
\end{equation}
where ${\rm{IFT(\cdot)}}$ denotes the inverse Fourier transform operation, ${{\rm{W}}_{{\rm{hann}}}}$ represents the Hanning-window. Then, the power delay profile (PDP) is calculated by 
\begin{equation}
\label{eq8}
{PDP} = {\left| {h(t,\tau )} \right|^2}.
\end{equation}

In this paper, the multi-path components (MPCs) of PDP above the detection threshold $P_{th}$ are considered valid, and otherwise the MPCs are deemed outliers and discarded. The detection threshold $P_{th}$ can be determined by \eqref{eq9} \cite{bs16}.
\begin{equation}
\label{eq9}
{P_{th}} = \max ({P_{\max}} - {\gamma _P},{N_0} + {\gamma_N}),
\end{equation}
where $P_{\max}$ is the maximum peak power of PDP, $\gamma_P$ is the power threshold relative to the maximum peak, $N_0$ is the noise floor of PDP, $\gamma_N$ is the power threshold relative to the noise floor. In this paper, $\gamma_P$ and $\gamma_N$ are set to $60$ dB and $15$ dB, respectively. 

For describing the channel shape in temporal domain, RMS DS can be represented as 
\begin{equation}
\label{eq11}
{\tau _{{\rm{RMS}}}}{\rm{ =  }}\sqrt {\frac{{\sum\nolimits_{l = 1}^L {{P_l}{\tau _l}^2} }}{{\sum\nolimits_{l = 1}^L {{P_l}} }} - {{\left( {\frac{{\sum\nolimits_{l = 1}^L {{P_l}{\tau _l}} }}{{\sum\nolimits_{l = 1}^L {{P_l}} }}} \right)}^2}} ,
\end{equation}
where $L$ is the number of the valid MPCs in PDP, $\tau_l$ denotes the delay of the $l$th valid MPC calculated by ${\tau _l} = (l - 1){\Delta _\tau}$, $\Delta _\tau$ denotes the delay resolution equaling to the inverse of the measured signal bandwidth, and $P_l$ denotes the power of the $l$th valid MPC. 

\vspace{-0.2cm}
\subsection{Channel modeling}

In this subsection, the modeling methods for RIS-assisted channels are investigated. In the existing channel-related work, the FI model has been adopted widely to describe the characteristics of PL in the 3rd Generation Partnership Project (3GPP) and International Telecommunication Union (ITU) standards \cite{bs20, bs21}. Meanwhile, though the CI model only focuses on the PLE, it still attracts great attention due to its model parameter stability and intuitive perspective. In \cite{bs15}, we have proposed the applicable FI and CI models to RIS-assisted channel. Nevertheless, these two models in \cite{bs15} only considered the single-dimensional variable, i.e., the Rx-RIS distance, which limited their generalizations applied in various scenarios. Thus, in this paper, we further refine them from single-dimensional variable to multi-dimensional variables. The modified FI and CI models are proposed for general RIS-assisted channels, which include distance information and angle information.

For the convenience of presentation, the traditional FI model is provided as follows.
\begin{equation}
\label{eq12}
PL_{FI} = \alpha  + 10\beta {\log _{10}}\left( d \right) + X_\sigma ^{FI},
\end{equation}
where $d$ (m) is the three-dimensional (3D) direct distance between Tx and Rx, $\beta$ is the PLE associated with the variation of distance, $\alpha$ is intercept parameter associated with the offset value of PL, $X_\sigma ^{FI}$ is shadow factor (SF) for describing the fitting deviation, which is usually modeled as Gaussian distribution $\mathcal{N}(\mu,\sigma^2)$ \cite{bs16}.

The traditional CI model can be expressed as 
\begin{equation}
\label{eq13}
PL_{CI} = P{L_{FS}}\left( {{d_0}} \right){\rm{ + 10}}n\log_{10}\left( {\frac{d}{{{d_0}}}} \right) + X_\sigma ^{CI},
\end{equation}
where $n$ is the PLE associated with the variation of distance $d$, $d_0$ is the reference distance, which is usually set to $1$ m, $P{L_{FS}}\left( {{d_0}} \right)$ is the PL at reference distance $d_0$ in free space, and $X_\sigma ^{CI}$ is SF following Gaussian distribution $\mathcal{N}(\mu,\sigma^2)$.

As we can see, both of the traditional FI and CI models focus on the PL fittings under one-dimensional variable (i.e., $d$). However, for RIS-assisted channel, there are multi-dimensional variables, e.g., $d_1$, $d_2$, $\theta_t$, $\theta_r$, etc., to be involved in channel modeling. The theoretical RIS-assisted channel models in free space were proposed in \cite{b5}, which can be expressed as
\begin{equation}
\label{eq14}
PL_{FS}^{RIS} = \frac{{16{\pi ^2}{{({d_1}{d_2})}^2}}}{{{G_t}{G_r}{{(MN{d_x}{d_y})}^2}\cos ({\theta _t})\cos ({\theta _r}){A^2}}},
\end{equation}
where $\cos ({\theta _t})$ and $\cos ({\theta _r})$ denote the normalized power radiation patterns of emission and reception for unit cell on RIS \cite{b5}. From this equation, it can be readily derived that, in free space, the PLEs on $d_1$ and $d_2$ are $2$ and the PLEs on $\cos ({\theta _t})$ and $\cos ({\theta _r})$ are $1$ in logarithmic scale, respectively. As a consequence, the traditional FI and CI models should be deformed to adapt to multi-dimensional variables of RIS-assisted channel.

In this paper, we propose the modified FI and CI models for the PL of RIS-assisted channel, as expressed in \eqref{eq15} and \eqref{eq16} respectively.
\begin{equation}
\label{eq15}
\begin{aligned}
PL_{FI}^{RIS}\left( {{d_1},{d_2},{\theta _t},{\theta _r}} \right) = 
& \alpha  + 10{\beta _1}{\log _{10}}\left( {{d_1}} \right) + 10{\beta _2}{\log _{10}}\left( {{d_2}} \right) - 10{\lambda _1}{\log _{10}}\left( {\cos \left( {{\theta _t}} \right)} \right)\\
& - 10{\lambda _2}{\log _{10}}\left( {\cos \left( {{\theta _r}} \right)} \right) + X_\sigma ^{FI},
\end{aligned}
\end{equation}
where $\beta _1$ and $\beta _2$ are the PLEs to describe the dependence of PL on $d_1$ and $d_2$, $\lambda_1$ and $\lambda_2$ are the PLEs to indicate the dependence of PL on $\theta_t$ and $\theta_r$, $\alpha$ denotes the intercept parameter associated with the offset value of PL. $X_\sigma ^{FI}$ denotes SF, which is modeled as Gaussian distribution $\mathcal{N}(\mu,\sigma^2)$.
\begin{equation}
\label{eq16}
\begin{aligned}
PL_{CI}^{RIS}\left( {{d_1},{d_2},{\theta _t},{\theta _r}} \right) = 
& PL_{FS}^{RIS}{\rm{(}}d_0^1,d_0^2{\rm{,}}\theta _0^t,\theta _0^r{\rm{)}} + 10{n_1}{\log _{10}}\left( {\frac{{{d_1}}}{{d_0^1}}} \right) + 10{n_2}{\log _{10}}\left( {\frac{{{d_2}}}{{d_0^2}}} \right) \\
&- 10{\mu _1}{\log _{10}}\left( {\frac{{\cos \left( {{\theta _t}} \right)}}{{\cos \left( {\theta _0^t} \right)}}} \right) - 10{\mu _2}{\log _{10}}\left( {\frac{{\cos \left( {{\theta _r}} \right)}}{{\cos \left( {\theta _0^r} \right)}}} \right) + X_\sigma ^{CI},
\end{aligned}
\end{equation}
where $n_1$ and $n_2$ are the PLEs to describe the dependence of PL on $d_1$ and $d_2$, $\mu_1$ and $\mu_2$ are the PLEs to indicate the dependence of PL on $\theta_t$ and $\theta_r$, $X_\sigma ^{CI}$ is SF following the Gaussian distribution $\mathcal{N}(\mu,\sigma^2)$, $d_0^1$, $d_0^2$, $\theta_0^t$, and $\theta _0^r$ denote the reference distances of $d_1$ and $d_2$ as well as the reference angles of $\theta_t$ and $\theta_r$, respectively. In this paper, $d_0^1$, $d_0^2$, $\theta_0^t$, and $\theta _0^r$ are set to $1$ m, $1$ m, $0^\circ$, and $0^\circ$ respectively, whereas these reference values can be adjusted appropriately in each scenario where applicable. Moreover, $PL_{FS}^{RIS}\left( {d_0^1,d_0^2{\rm{,}}\theta _0^t,\theta _0^r} \right)$ represents the PL with these reference variables in free space, which is calculated to be $21.38$ dB according to \eqref{eq14}. Then, a simplified expression of \eqref{eq16} can be written as  
\begin{equation}
\label{eq17}
\begin{aligned}
PL_{CI}^{RIS}\left( {{d_1},{d_2},{\theta _t},{\theta _r}} \right) = & 21.38 + 10{n_1}{\log _{10}}\left( {{d_1}} \right) + 10{n_2}{\log _{10}}\left( {{d_2}} \right) - 10{\mu _1}{\log _{10}}\left( {\cos \left( {{\theta _t}} \right)} \right) \\
& - 10{\mu _2}{\log _{10}}\left( {\cos \left( {{\theta _r}} \right)} \right) + X_\sigma ^{CI}.
\end{aligned}
\end{equation}

In this paper, the least square (LS) method is utilized to fit the measurement data with the modified FI and CI models, respectively. In detail, LS method adopts Levenberg-marquardt algorithm, which requires presetting the upper and lower bounds of the parameters to be fitted. In this paper, the upper and lower bounds are set to $(50, 3, 3, 2, 2)$ and $(10, 1, 1, 0, 0)$ respectively, for the parameters ($\alpha$, $\beta_1$, $\beta_2$, $\lambda_1$, $\lambda_2$) in the modified FI model. Moreover, the upper and lower bounds are set to $(21.38, 3, 3, 2, 2)$ and $(21.38, 1, 1, 0, 0)$ respectively, for the parameters ($PL_{FS}^{RIS}$, $n_1$, $n_2$, $\mu_1$, $\mu_2$) in the modified CI model.

\section{Measurement Results and Modeling Analyses}
\label{sec5}
In this section, the measured channel data in three measurement campaigns is presented and compared with the free-space channel model proposed in \cite{b5}. The modified CI and FI models are fitted with the measured channel data under \textit{intelligent reflection with RIS}. Moreover, the channel characteristics in temporal, frequency, and spatial domains including PDP, RMS DS, frequency stationarity, spatial consistency, etc., are analyzed and explained.

\vspace{-0.2cm}
\subsection{Outdoor measurement}

For the mode of intelligent reflection with RIS in outdoor measurement, the fitting results of the modified CI and FI models with measurement data as well as free-space data are summarized in \autoref{table5}. From \autoref{table5}, it can be found that for both of the modified CI and FI models, the fitted PLE on $d_1$ with free-space data is slightly lower than that with measurement data. On the contrary, the fitted PLE on $d_2$ with free-space data is slightly higher than that with measurement data. In addition, similar phenomena can be seen in angle domain that the fitted PLEs on $\theta_t$ and on $\theta_r$ with free-space data are respectively lower and higher than those with measurement data. These phenomena may result from that in outdoor measurement, Rx positions are surrounded by abundant scatterers such as neighbouring buildings while Tx positions are next to open streets.

\begin{table}[htbp]
\vspace{-0.5cm}
\centering
\footnotesize
\caption{Fitting results for outdoor measurement.}
\label{table5}
 \vspace{-0.2cm}
\begin{tabular}{|c|c|c|c|c|c|}
\hline
Parameters & $\alpha$ ($PL_{FS}^{RIS}$) & $\beta_1$ ($n_1$)
& $\beta_2$  ($n_2$) & $\lambda_1$  ($\mu_1$) & $\lambda_2$ ($\mu_2$) \\ \hline
Modified FI with \textbf{measurement} data & $20.08$ & $2.29$
& $1.88$ & $1.21$ & $0.65$ \\\hline
Modified FI with \textbf{free-space} data & $24.52$ & $2.05$
& $1.95$ & $1.15$ & $0.79$ \\\hline
Modified CI with \textbf{measurement} data & $21.38$ & $2.22$
& $1.82$ & $1.21$ & $0.64$ \\\hline
Modified CI with \textbf{free-space} data & $21.38$ & $2.21$ & $2.08$ & $1.14$ & $0.81$ \\\hline
\end{tabular}
\vspace{-0.5cm}
\end{table}

\begin{figure}[htbp]
\vspace{-0.5cm}
\centering
\begin{minipage}[t]{0.48\textwidth}
\centering
\includegraphics[width=3in]{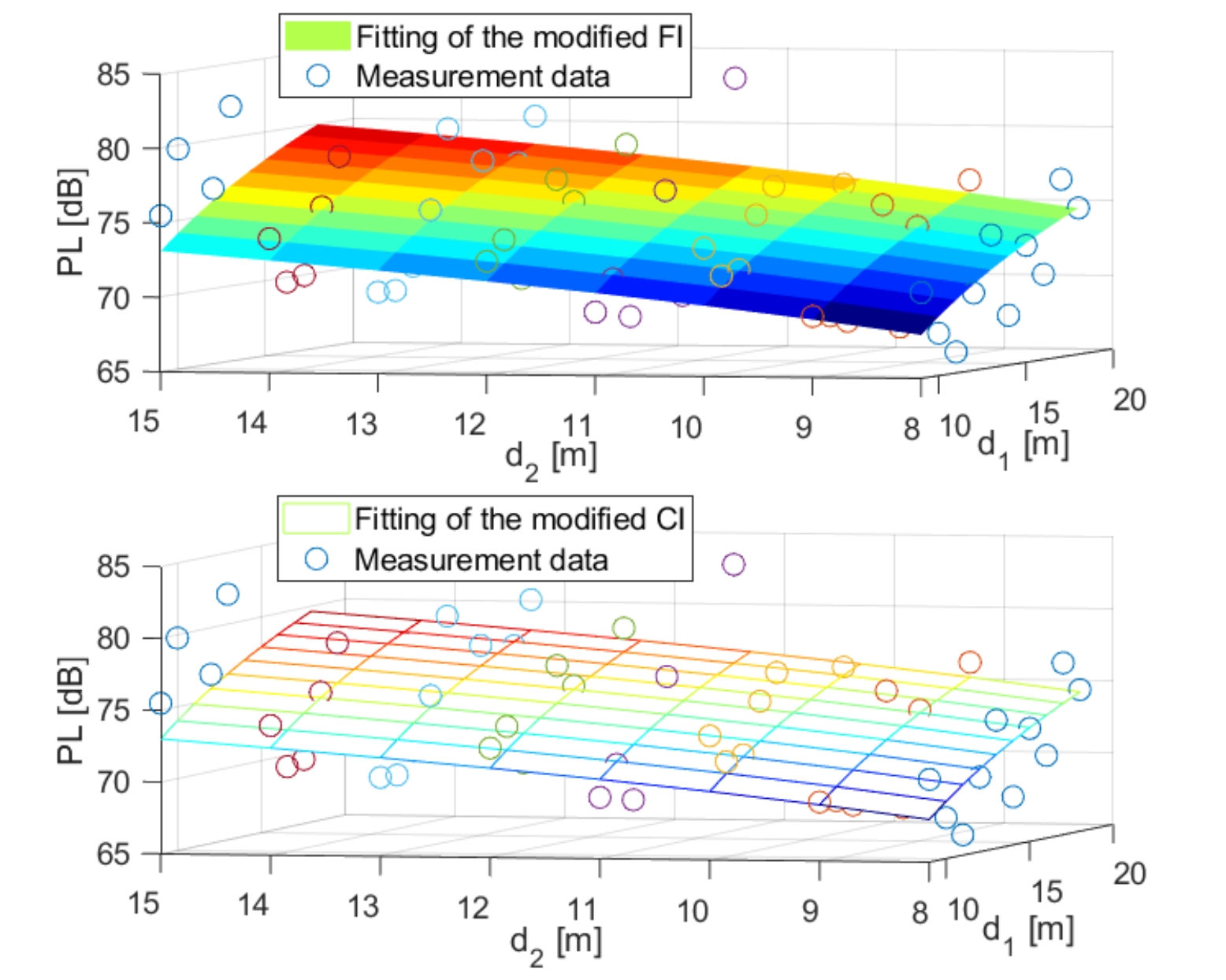}
\vspace{-0.5cm}
\caption{The modified models with measurement data.}
\label{fig9}
\end{minipage}
\begin{minipage}[t]{0.48\textwidth}
\centering
\includegraphics[width=3in]{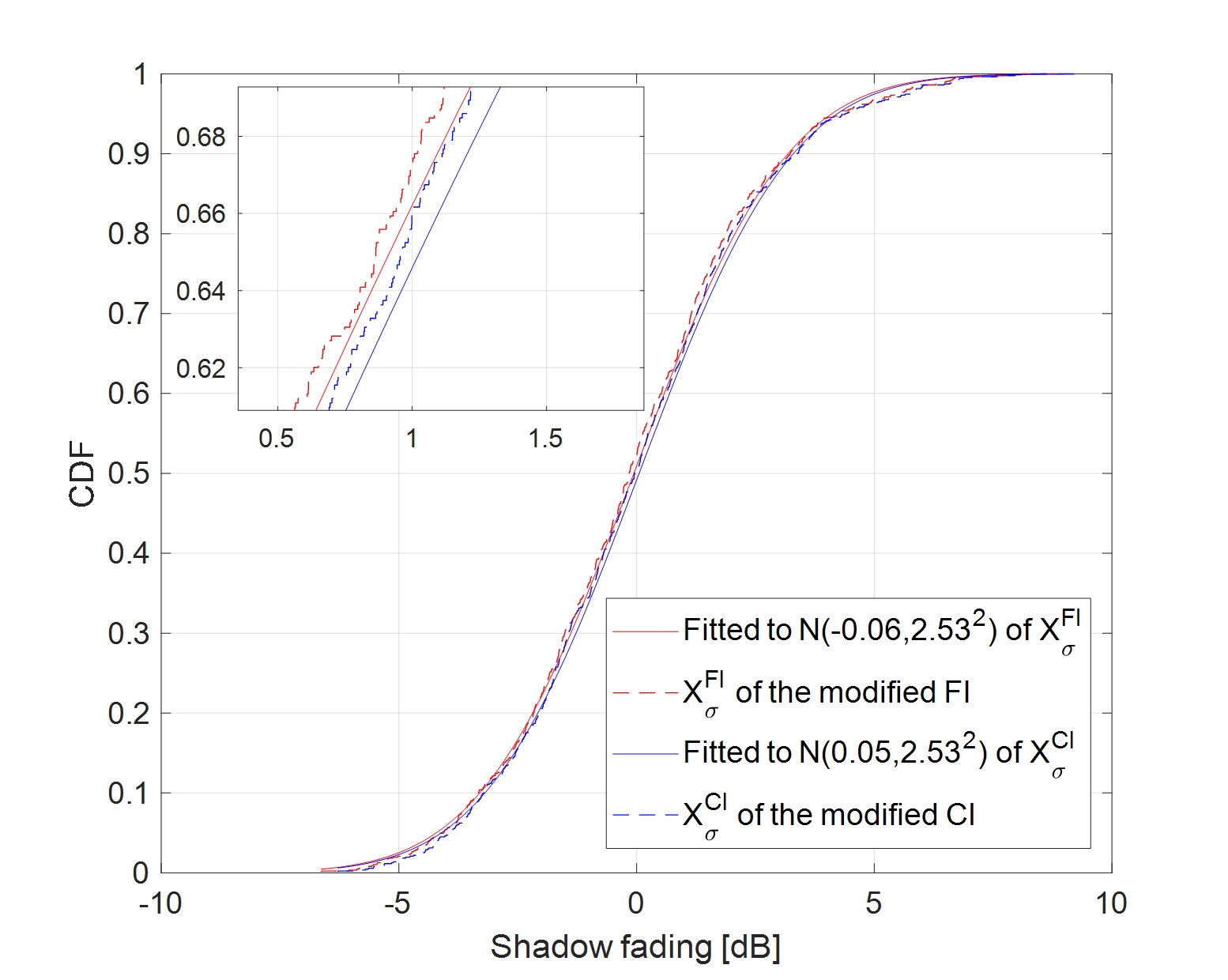}
\vspace{-0.5cm}
\caption{SFs and the fitted CDFs.}
\label{fig10}
\end{minipage}
\vspace{-0.5cm}
\end{figure}

\reffig{fig9} depicts the modified CI and FI models fitted with the measurement data under $\theta_t = 75^\circ$ and $\theta_r = 60^\circ$. It can be visually observed that both of the proposed models fit well with the measurement data. \reffig{fig10} exhibits the cumulative distribution functions (CDFs) of SFs for the modified FI and CI models, which are fitted into the Gaussian distributions with a same standard deviation of $\sigma = 2.53$. It delivers a similar prediction accuracy of the modified FI and CI models considering their similar standard deviation.

\begin{figure}[htbp]
 \vspace{-0.5cm}
\centering
\begin{minipage}[t]{0.48\textwidth}
\centering
\includegraphics[width=3in]{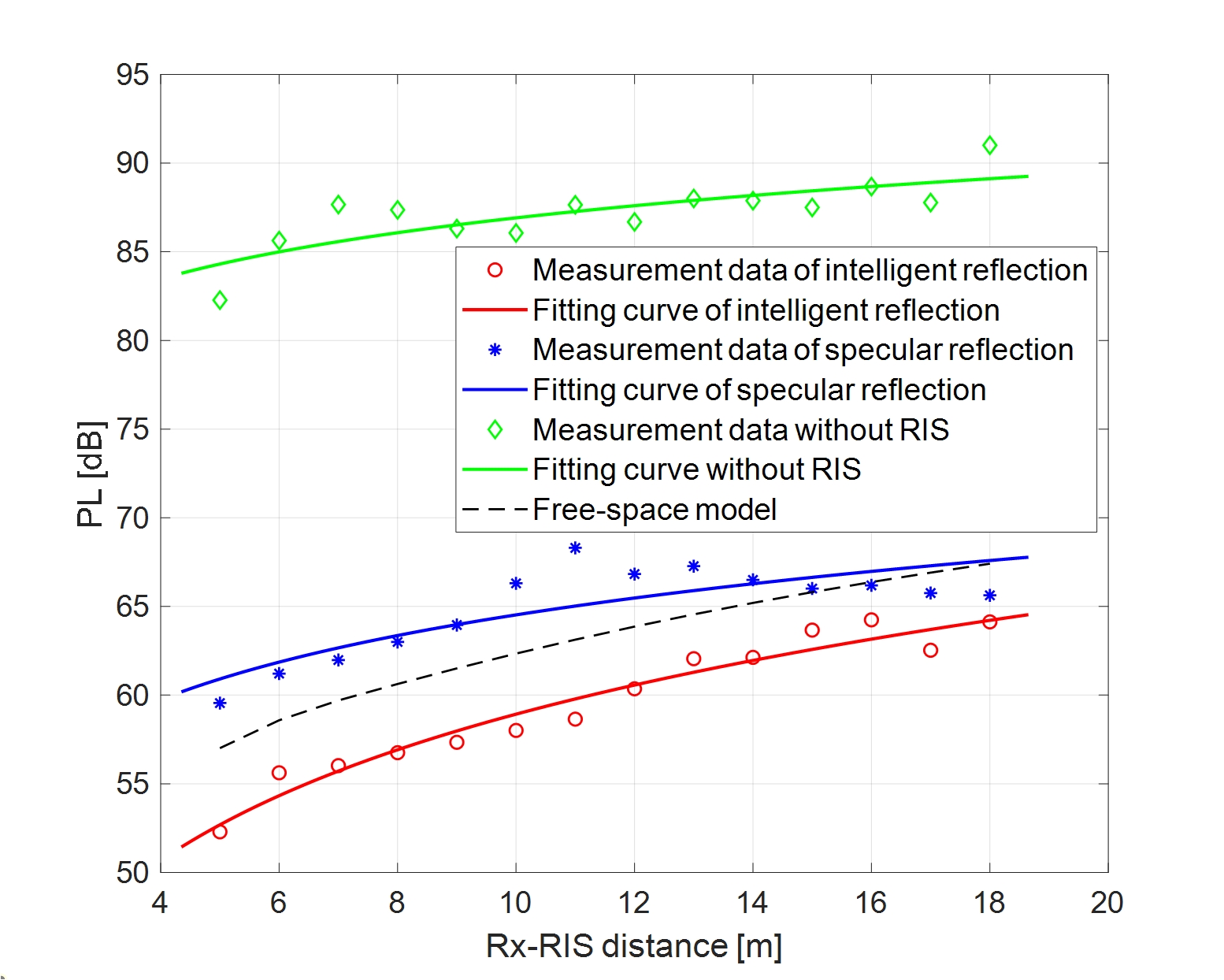}
\vspace{-0.5cm}
\caption{PLs of measurement data and free-space data.}
\label{fig11}
\end{minipage}
\begin{minipage}[t]{0.48\textwidth}
\centering
\includegraphics[height=2.6in]{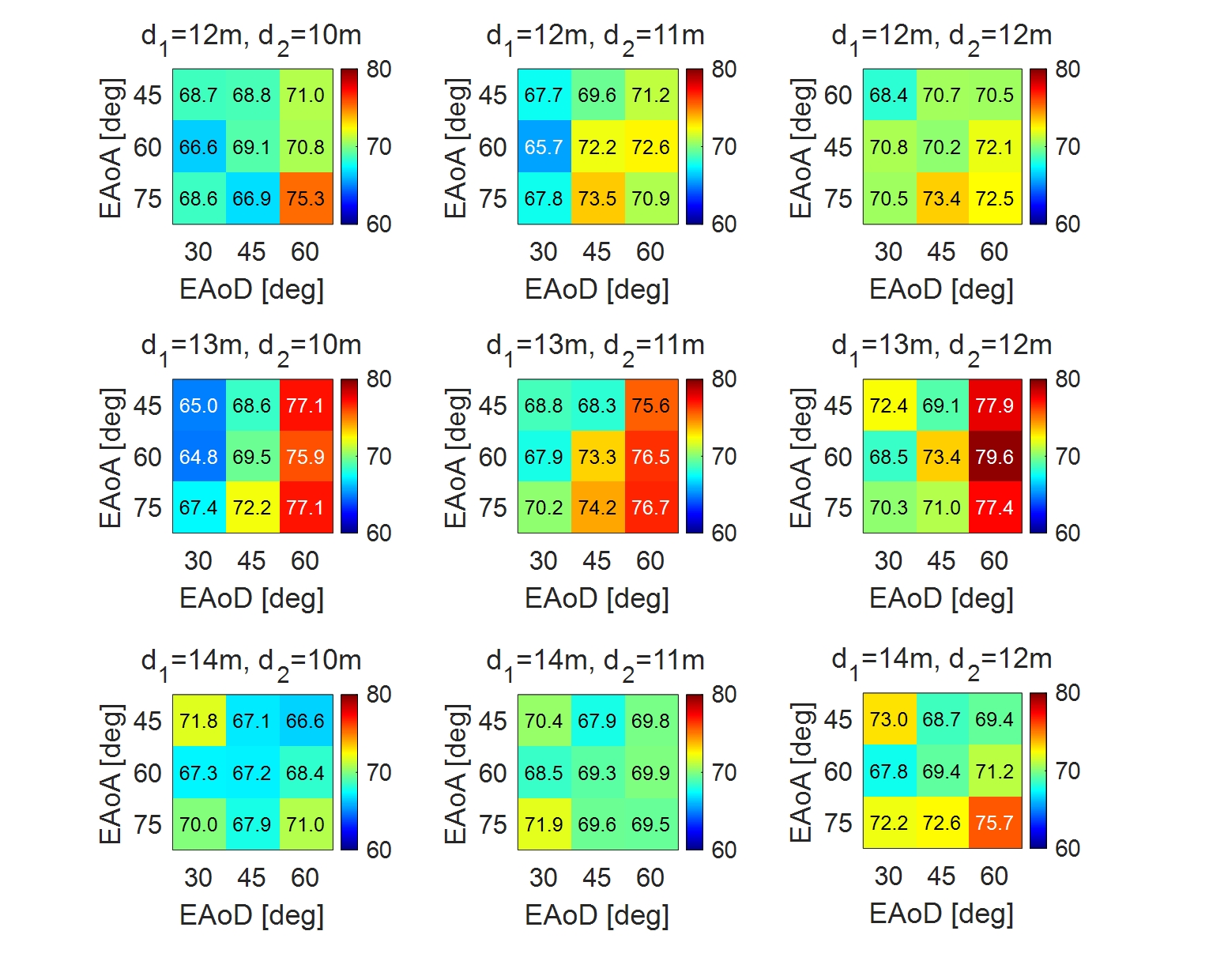}
\vspace{-0.5cm}
\caption{PLs under $d_1=12\sim14$ m and $d_2=10\sim12$ m.}
\label{fig12}
\end{minipage}
\vspace{-0.5cm}
\end{figure}

\reffig{fig11} provides the measured PL for three modes in outdoor scenario as well as the free-space PL, under $\theta_t = 45^\circ$, $\theta_r = 45^\circ$, and $d_1 = 6$ m. From this figure, the PLE for intelligent reflection mode shows a similar value as the free-space model, though there is a small gap of about $3$ dB between them. In addition, thanks to the capability of focusing signal energy for intelligent reflection mode, its PL demonstrates a maximum gain of $9.7$ dB than that for specular reflection mode and a maximum gain of $30$ dB than that without RIS, respectively. As $d_2$ increases, the PL difference between intelligent reflection and specular reflection becomes smaller. When $d_2$ is large enough, their PLs tend to coincide. This shows that specular reflection and intelligent reflection behave equally in the case of extremely far field, when $\theta_t=\theta_r$. \reffig{fig12} illustrates the PLs vs EAoA and EAoD under $d_1 = 12\sim14$ m and $d_2 = 10\sim12$ m for intelligent reflection mode, with color coded against PL. From this figure, the PL keeps up the trend of increasing as EAoA and EAoD become larger, despite a slight fluctuation. This phenomenon is well consistent with the theoretical PL scaling law in angle domain described in \cite{b4, b5}.

\begin{figure}[htbp]
\vspace{-0.5cm}
\centering
\begin{minipage}[t]{0.48\textwidth}
\centering
\includegraphics[height=2.4in]{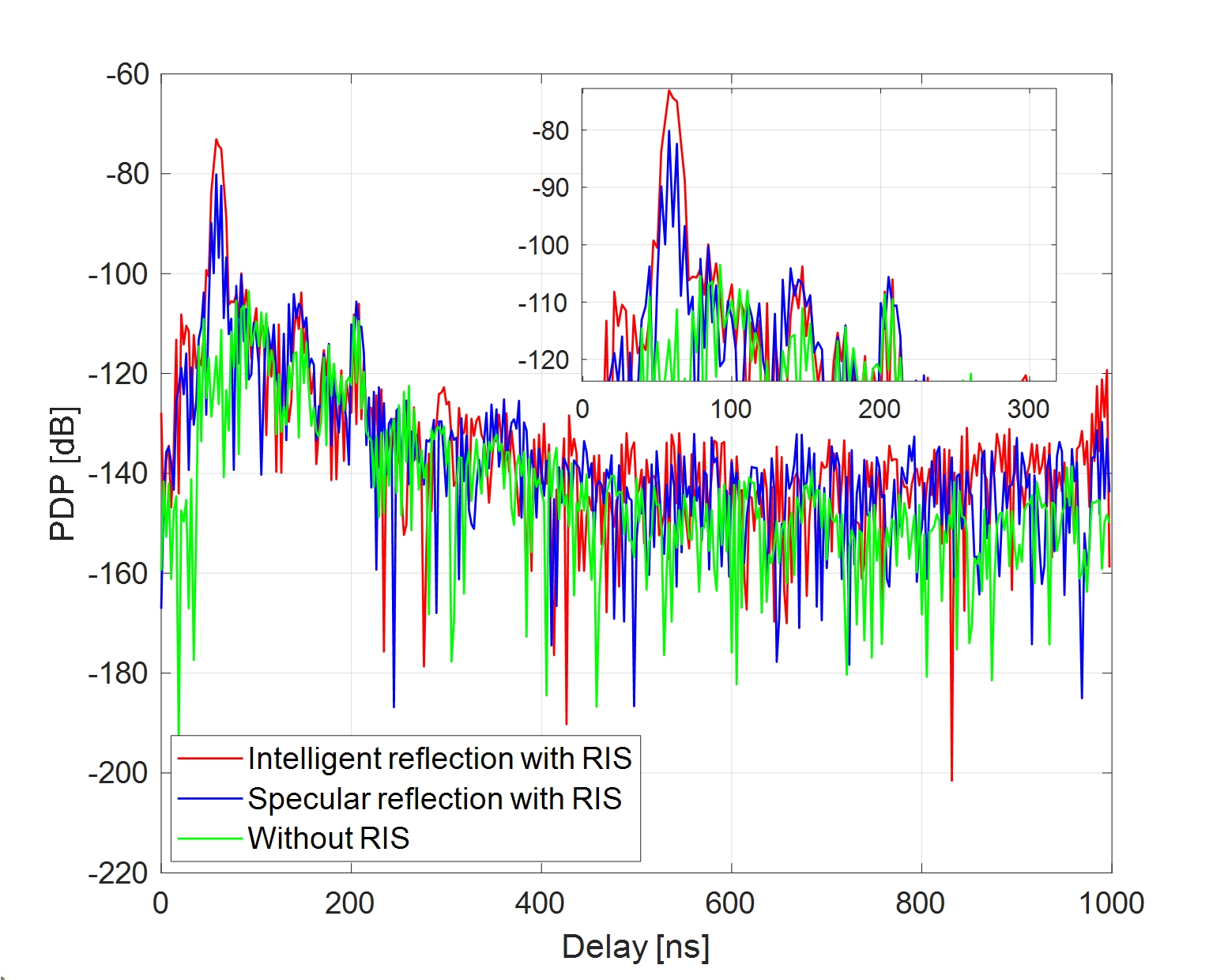}
\vspace{-0.5cm}
\caption{PDP comparison of three modes.}
\label{fig13}
\end{minipage}
\begin{minipage}[t]{0.48\textwidth}
\centering
\includegraphics[height=2.4in]{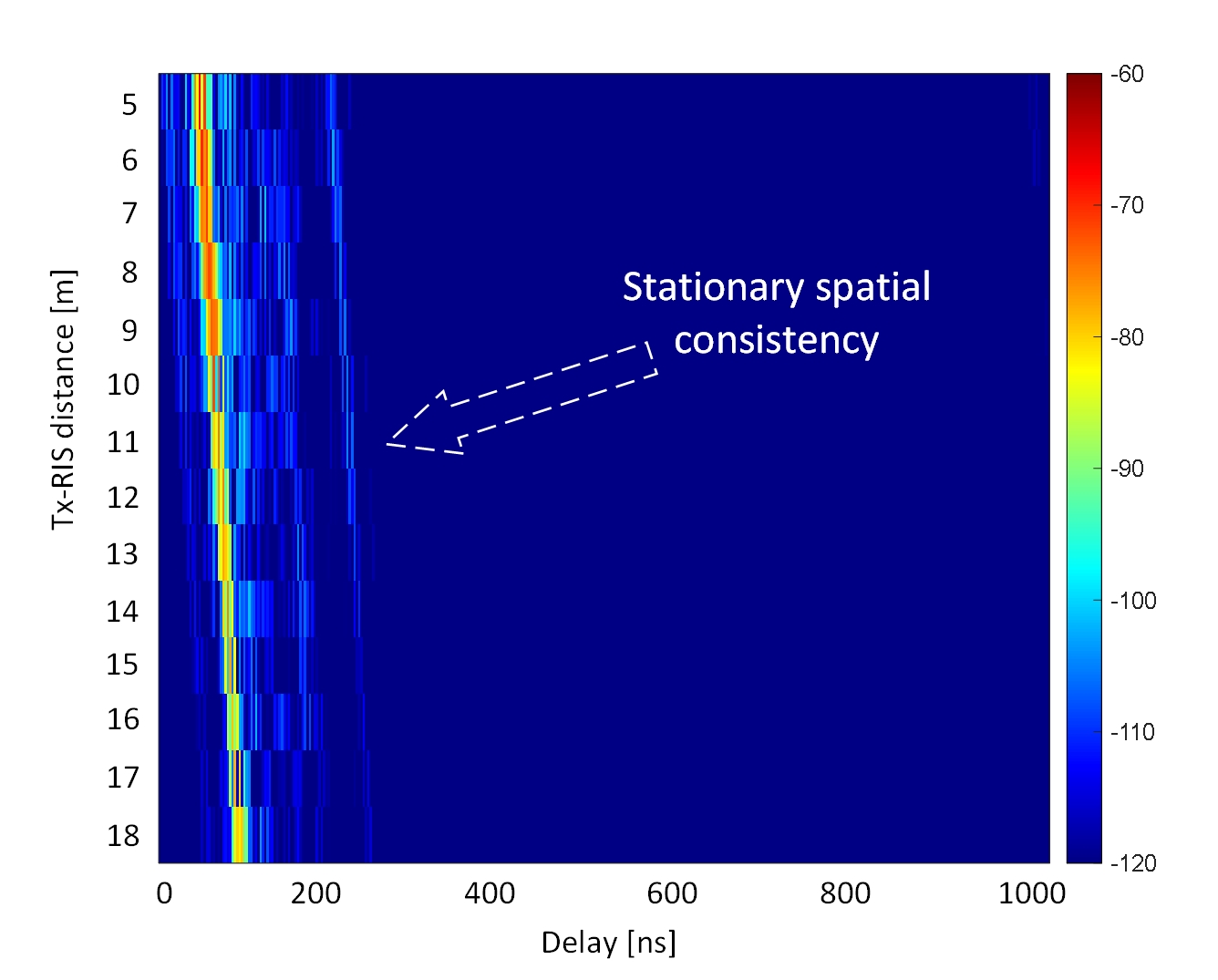}
\vspace{-0.5cm}
\caption{PDP evolution vs delay and position.}
\label{fig14}
\end{minipage}
\vspace{-0.2cm}
\end{figure}

\reffig{fig13} demonstrates the PDP comparison of three modes under $\theta_t = 45^\circ$, $\theta_r = 45^\circ$, $d_1 = 9$ m, and $d_2 = 8$ m. From this figure, the PDP peak of intelligent reflection mode is highest, followed by that of specular reflection mode, and the PDP peak without RIS is lowest due to its inherent NLOS propagation. Moreover, the delays on PDP peaks of intelligent reflection mode and specular reflection mode are almost the same, which is lower than that without RIS. This can be explained by that the signal propagation without RIS is dominated by the scattering and diffraction aroused from the surroundings, which has a larger excess delay. From field-of-view of Rx, \reffig{fig14} illustrates the PDP evolution vs $d_1$ and delay for intelligent reflection mode, which are linked across positions, under $d_2 = 8$ m, $\theta_t = 45^\circ$, $\theta_r = 45^\circ$. It indicates a well stationary spatial consistency on MPC evolution when Tx is moved in our measurement.

\begin{figure}[htbp]
\vspace{-0.5cm}
\centering
\begin{minipage}[t]{0.48\textwidth}
\centering
\includegraphics[width=3in]{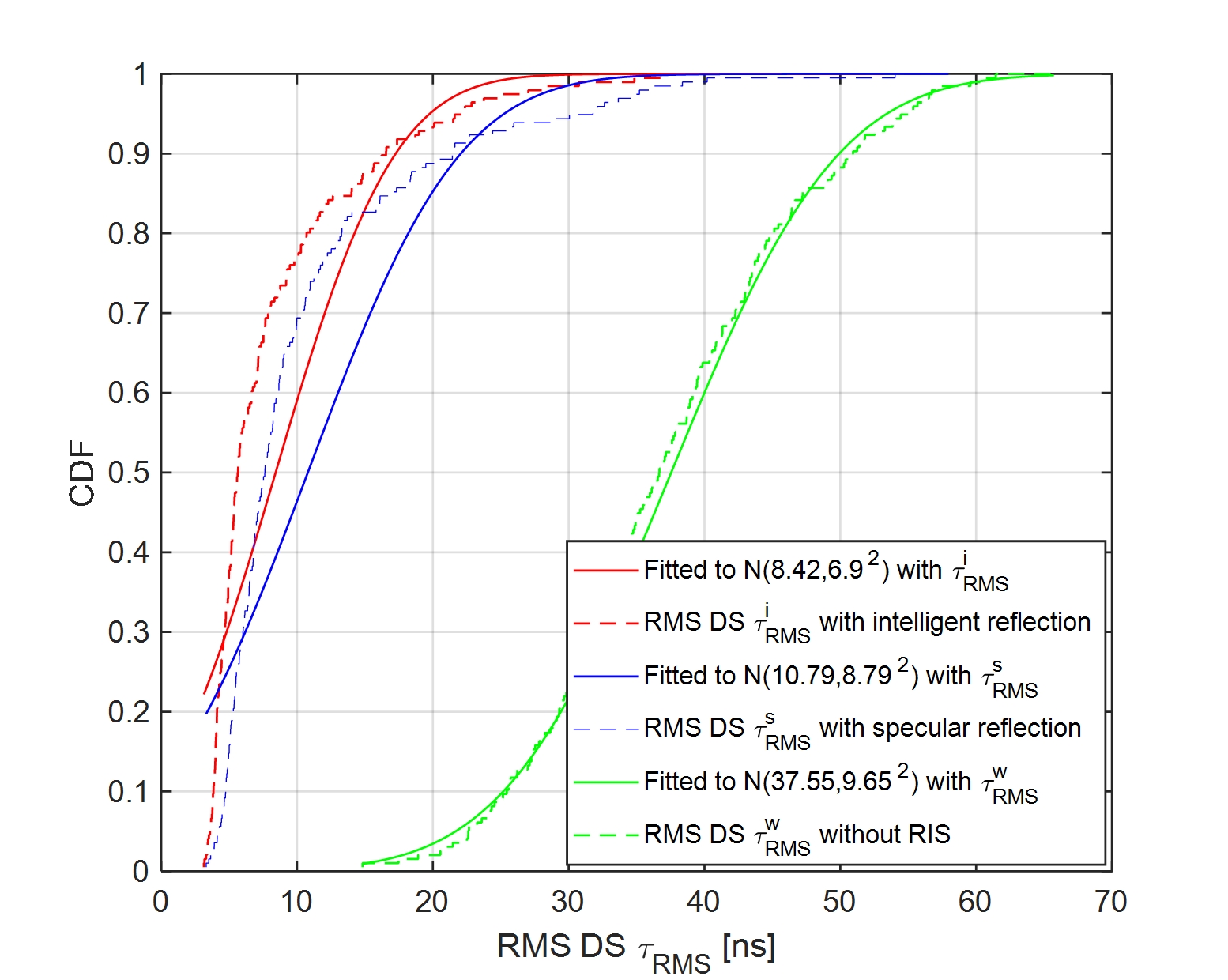}
\vspace{-0.5cm}
\caption{RMS DSs for three modes.}
\label{fig15}
\end{minipage}
\begin{minipage}[t]{0.48\textwidth}
\centering
\includegraphics[width=3in]{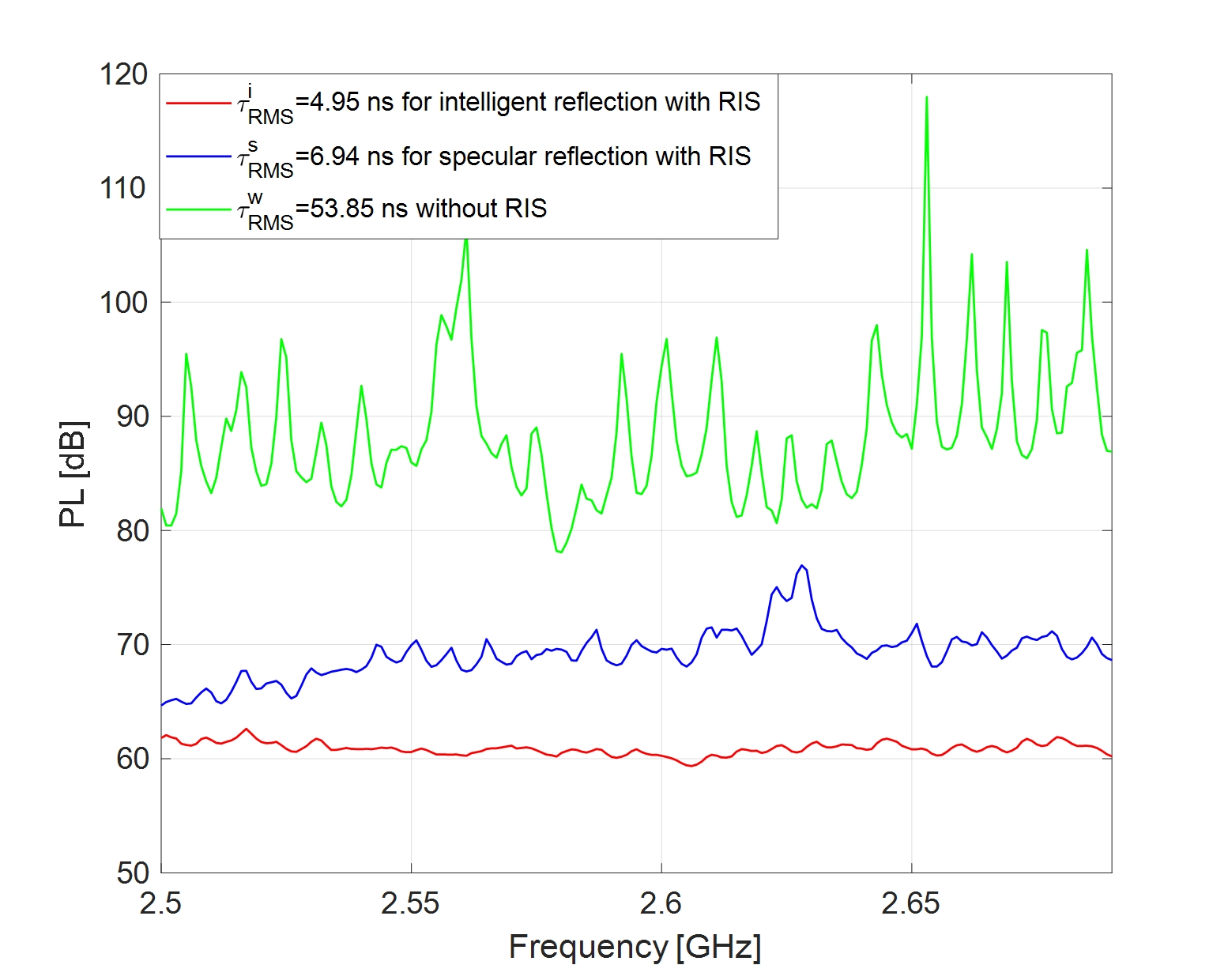}
\vspace{-0.5cm}
\caption{Frequency stationarity for three modes.}
\label{fig16}
\end{minipage}
\vspace{-0.5cm}
\end{figure}

\reffig{fig15} shows the RMS DSs for three modes and the fitted Gaussian distributions $\mathcal{N}(\mu,\sigma^2)$, under $\theta_t = 45^\circ$ and $\theta_r = 45^\circ$. The means of these CDFs are $8.42$ ns, $10.79$ ns and $37.55$ ns respectively for the modes of intelligent reflection, specular reflection and without RIS, as displayed in \autoref{table_v1}. The higher $\mu$ represents the more diffuse multipath distribution. This phenomenon illustrates that the channel time dispersion of intelligent reflection is lower than that of specular reflection, because the latter may cause diffuse reflection of EM waves. In addition, the channel time dispersion without RIS is strongest, on account of that it is dominated by scattering, diffraction, and diffuse reflection. 

\begin{table}[htbp]
 \vspace{-0.5cm}
\centering
\footnotesize
\caption{Fitted distributions of RMS DSs.}
\label{table_v1}
 \vspace{-0.2cm}
\begin{tabular}{|c|c|c|c|c|c|c|}
\hline
\multirow{2}{*}{\tabincell{c}{\textbf{Scenario}}} 
& \multicolumn{2}{|c|}{\textbf{Without RIS}}
& \multicolumn{2}{|c|}{\textbf{Specular reflection}}
& \multicolumn{2}{|c|}{\textbf{Intelligent reflection}} \\ 
\cline{2-7} 
& $\mu$ [ns] & $\sigma$ [ns] & $\mu$ [ns] & $\sigma$ [ns] & $\mu$ [ns] & $\sigma$ [ns] \\ \hline
\textbf{Outdoor} & $37.55$ & $9.65$ & $10.79$ & $8.79$ & $8.42$ & $6.9$ \\ \hline
\textbf{Indoor} & $13.04$ & $4.29$ & $5.88$ & $4.31$ & $5.01$ & $2.74$ \\ \hline
\textbf{O2I} & $24.21$ & $8.2$ & $16.04$ & $9.69$ & $10.36$ & $9.53$ \\ \hline
\end{tabular}
\vspace{-0.2cm}
\end{table}

\reffig{fig16} demonstrates the frequency stationarity within the measured bandwidth for three modes under $\theta_t = 45^\circ$, $\theta_r = 45^\circ$, $d_1 = 10$ m and $d_2 = 10$ m. From this figure, the propagation signal without RIS shows the strongest frequency selective fading, due to the fact that its dominant propagation originates from rich MPCs. In addition, the propagation signal for specular reflection with RIS demonstrates a slight but non-negligible frequency selective fading, which confirms that this mode still generates the additional MPCs when assisting desired signal. By contrast, the intelligent reflection with RIS has a nearly flat signal, illuminating its beamforming capability without creating excrescent MPCs. These phenomena can be numerically explained from their RMS DSs of $\tau _{\rm{RMS}}^{\rm{w}}=53.85$ ns, $\tau _{\rm{RMS}}^{\rm{s}}=6.94$ ns, and $\tau _{\rm{RMS}}^{\rm{i}}=4.95$ ns respectively, considering the signal coherence bandwidth $B_c$ is inversely proportional to its RMS DS, as \eqref{eq18}.
\begin{equation}
\label{eq18}
{B_c} \propto \frac{1}{{{\tau_{\rm{RMS}}}}}.
\end{equation}
\vspace{-0.5cm}

\vspace{-0.2cm}
\subsection{Indoor measurement}

\begin{figure}[htbp]
\vspace{-0.5cm}
\centering
\begin{minipage}[t]{0.48\textwidth}
\centering
\includegraphics[width=3in]{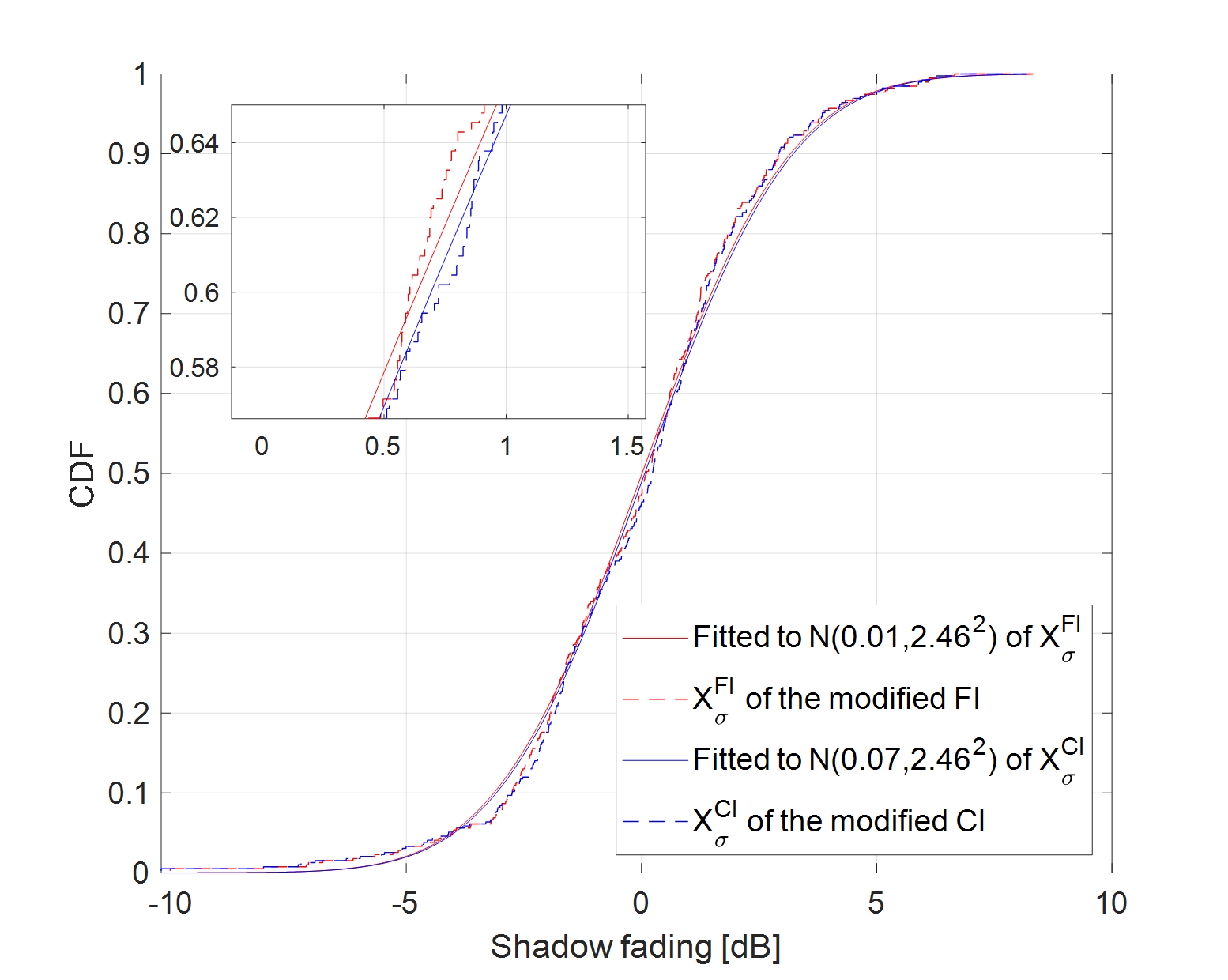}
\vspace{-0.5cm}
\caption{SFs and the fitted CDFs.}
\label{fig17}
\end{minipage}
\begin{minipage}[t]{0.48\textwidth}
\centering
\includegraphics[width=3in]{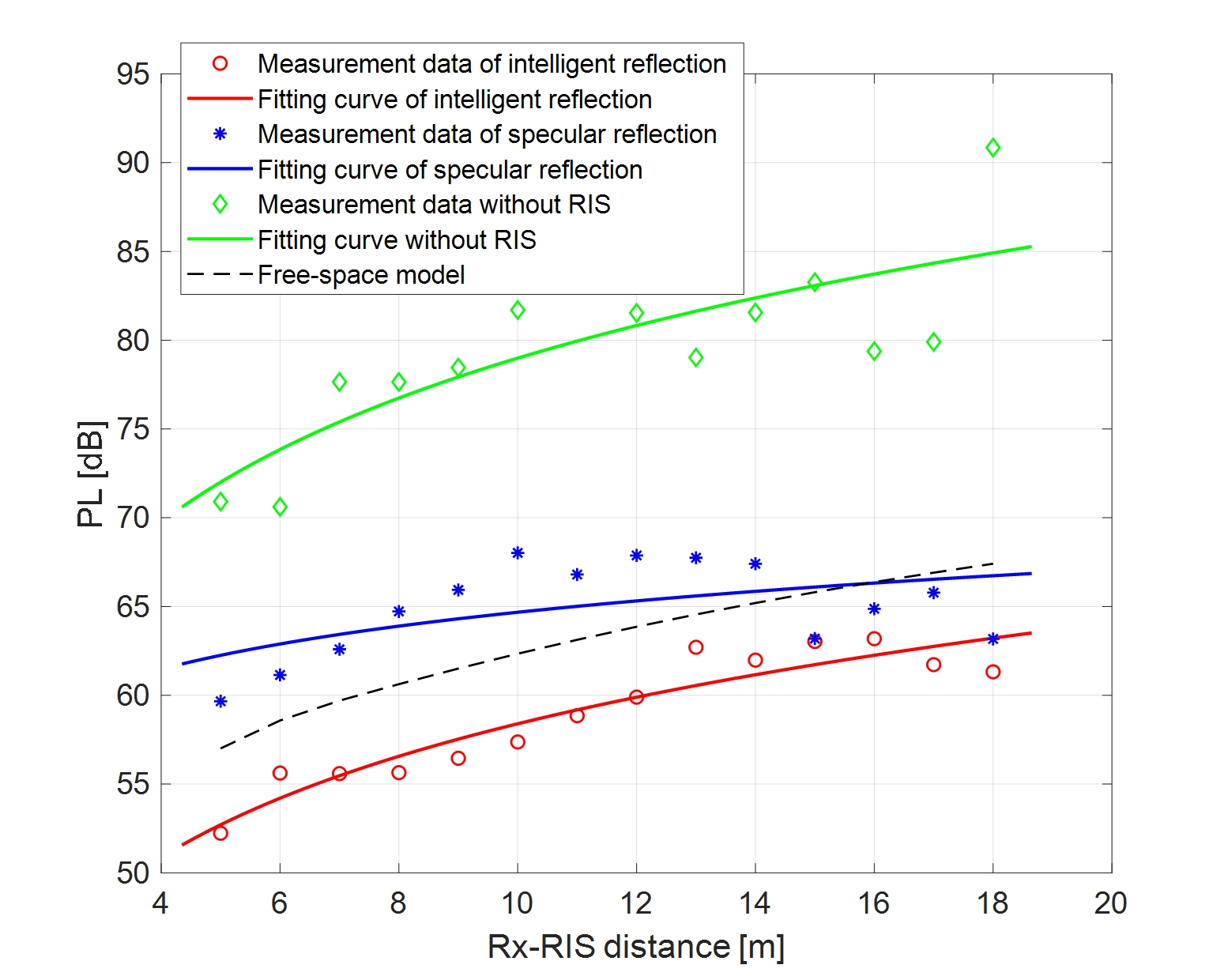}
\vspace{-0.5cm}
\caption{PLs of measurement data and free-space data.}
\label{fig18}
\end{minipage}
\vspace{-0.5cm}
\end{figure}

For the mode of intelligent reflection with RIS in indoor measurement, the fitting results of the modified CI and FI models with the measured data as well as the free-space data are summarized in \autoref{table6}. Note that in this measurement, the EAoA and EAoD are fixed to be $45^\circ$, thus $\lambda_1$ ($\mu_1$) and $\lambda_2$ ($\mu_2$) of the modified FI (CI) model are not taken into account. The reference value $PL_{FS}^{RIS} = 26.97$ of the modified CI model is calculated under $\theta_t = 45^\circ$, $\theta_r = 45^\circ$, $d_1 = 1$ m, and $d_2 = 1 $ m. From this table, it can be found that for both of the modified CI and FI models, the fitted PLEs on $d_1$ and $d_2$ with free-space data are slightly higher than those with measurement data. This can be explained by that in indoor corridor scenario, there exists a non-ignorable wave-guide effect, which results in the lower PLEs. \reffig{fig17} exhibits the CDFs of SFs for the modified models, both of which are fitted into the Gaussian distributions with $\sigma = 2.46$. Similar to the outdoor scenario, a similar prediction accuracy between the modified FI and CI models is observed for indoor corridor channels.

\reffig{fig18} shows the measured PLs for three modes as well as the free-space PL, under $\theta_t = 45^\circ$, $\theta_r = 45^\circ$, and $d_1 = 6$ m. From this figure, the PLE of intelligent reflection mode still shows a approximative value with the free-space PLE, though a small gap of about $4$ dB between them is observed. In addition, the PL for intelligent reflection mode demonstrates a maximum gain of $10.6$ dB than that for specular reflection mode and a maximum gain of $29.5$ dB than that without RIS. Similarly to \reffig{fig11}, as $d_2$ increases, the PL difference between intelligent reflection and specular reflection get smaller. When $d_2$ is large enough, their coinciding PLs can be inferred.

\begin{figure*}[htbp]
\vspace{-0.5cm}
     \centering
     \subfloat[]{\includegraphics[height = 1.75in]{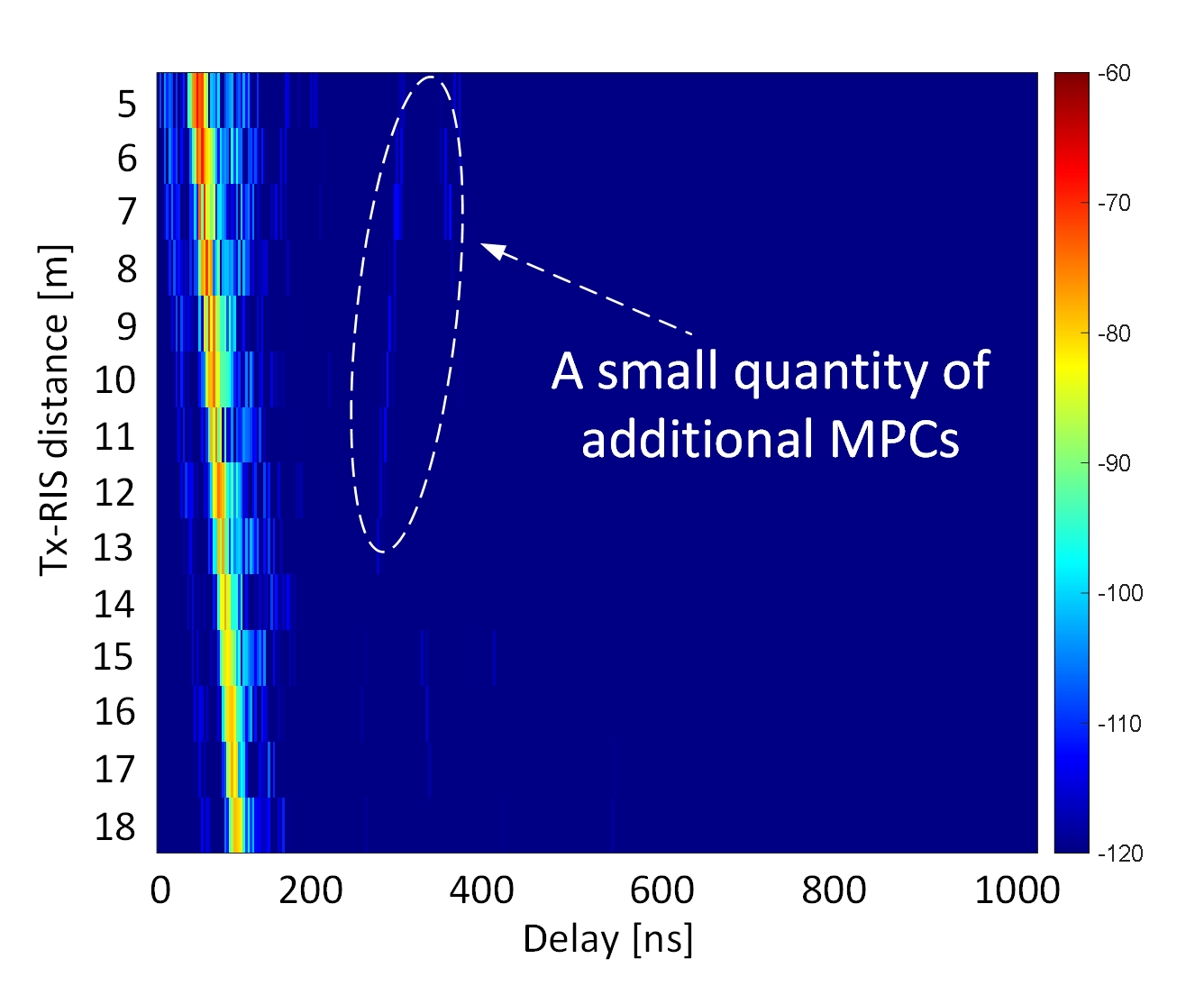}\label{fig20a}} 
     \subfloat[]{\includegraphics[height = 1.75in]{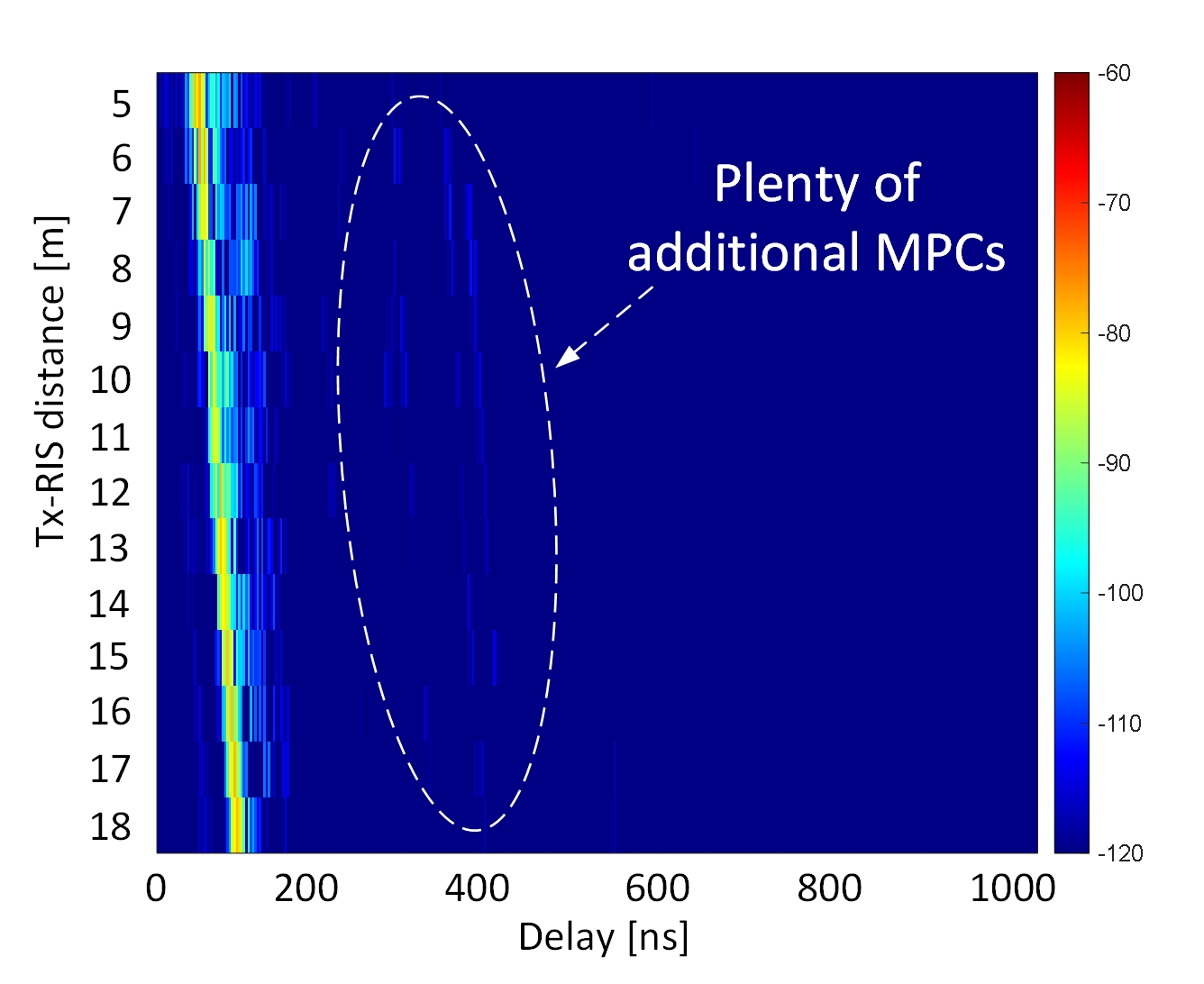}\label{fig20b}} 
     \subfloat[]{\includegraphics[height = 1.75in]{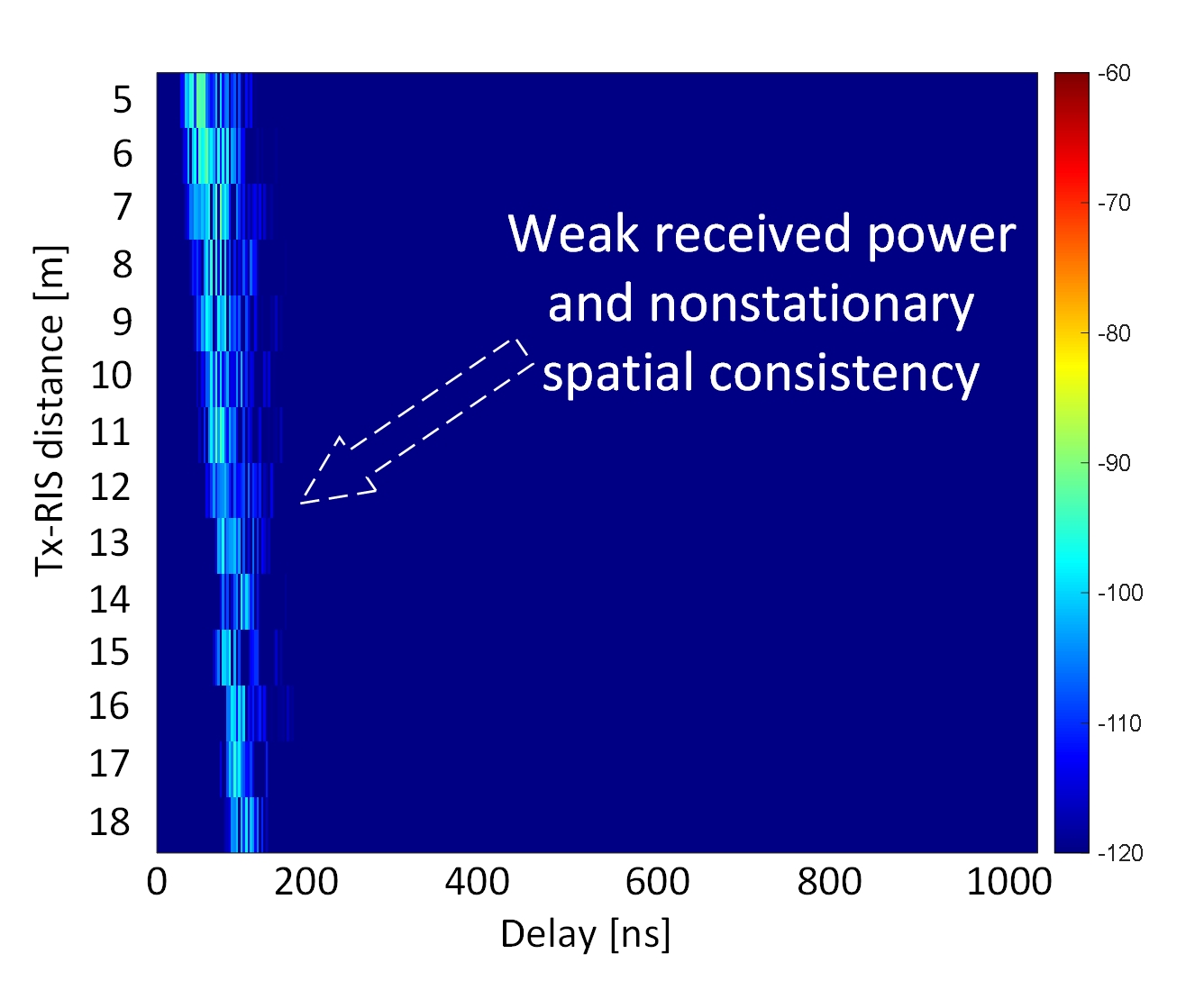}\label{fig20c}}
     \vspace{-0.2cm}
\caption{PDP evolution vs delay and position. (a) Intelligent reflection. (b) Specular reflection. (c) Without RIS.}
\label{fig20}
\vspace{-0.8cm}
\end{figure*}

\reffig{fig20} illustrates the PDP evolution vs $d_1$ and delay for three modes respectively, under $d_2 = 8$ m, $\theta_t = 45^\circ$, $\theta_r = 45^\circ$. Firstly, for the intelligent reflection mode, \reffig{fig20a} exhibits a stationary spatial consistency on MPC evolution, despite it shows a small quantity of additional MPCs at $300$ ns when $d_1$ is $5\sim13$ m. For the specular reflection mode in \reffig{fig20b}, there are plenty of additional MPCs covering all of the Tx positions. For the mode without RIS, a nonstationary spatial evolution with weak power is observed in \reffig{fig20c}. The above phenomena illustrate that the intelligent reflection mode of RIS is more conducive to focusing signal energy.

\begin{figure}[htbp]
\vspace{-0.5cm}
\centering
\begin{minipage}[t]{0.48\textwidth}
\centering
\includegraphics[width=3in]{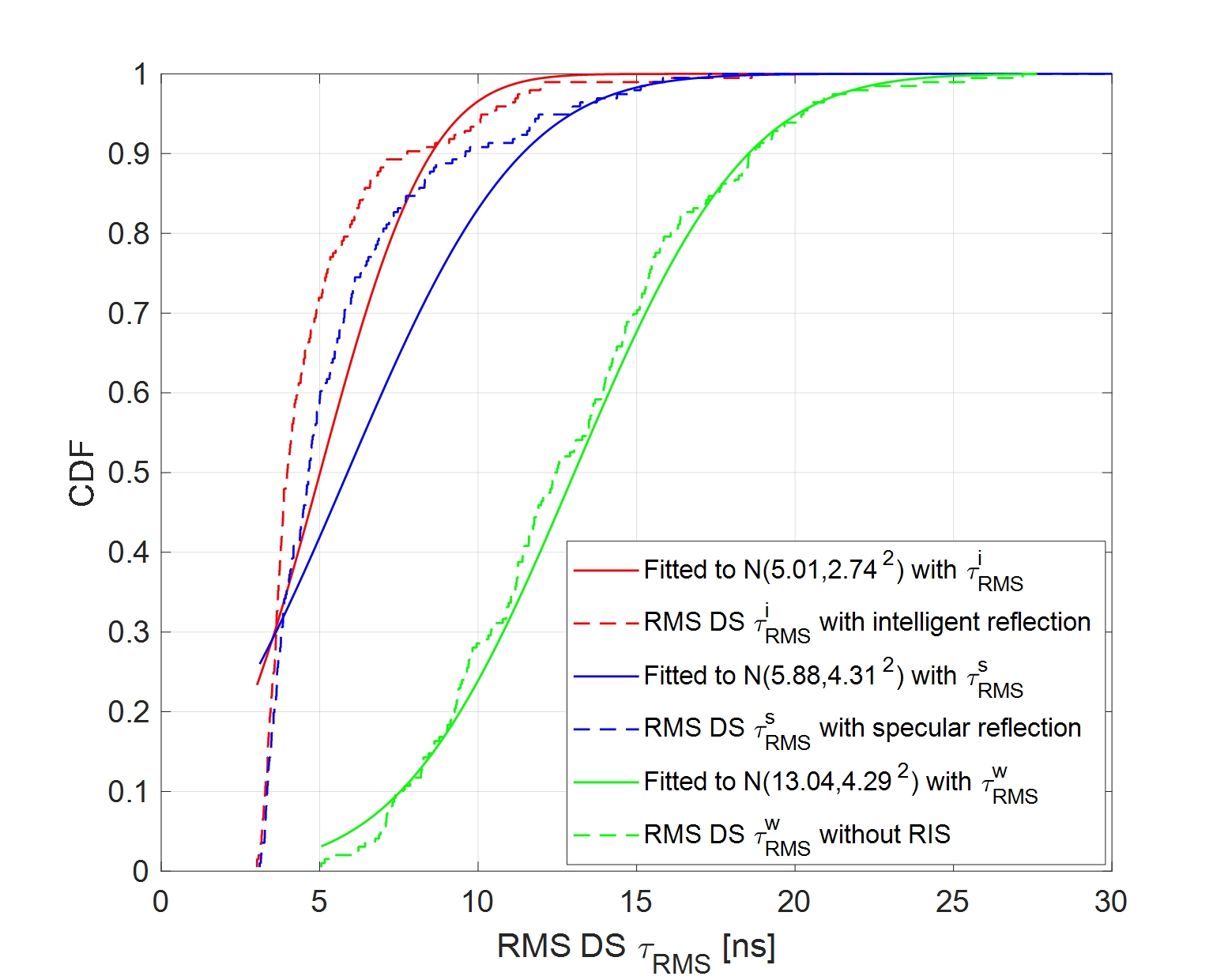}
\vspace{-0.5cm}
\caption{Comparison of RMS DSs for three modes.}
\label{fig21}
\end{minipage}
\begin{minipage}[t]{0.48\textwidth}
\centering
\includegraphics[width=3in]{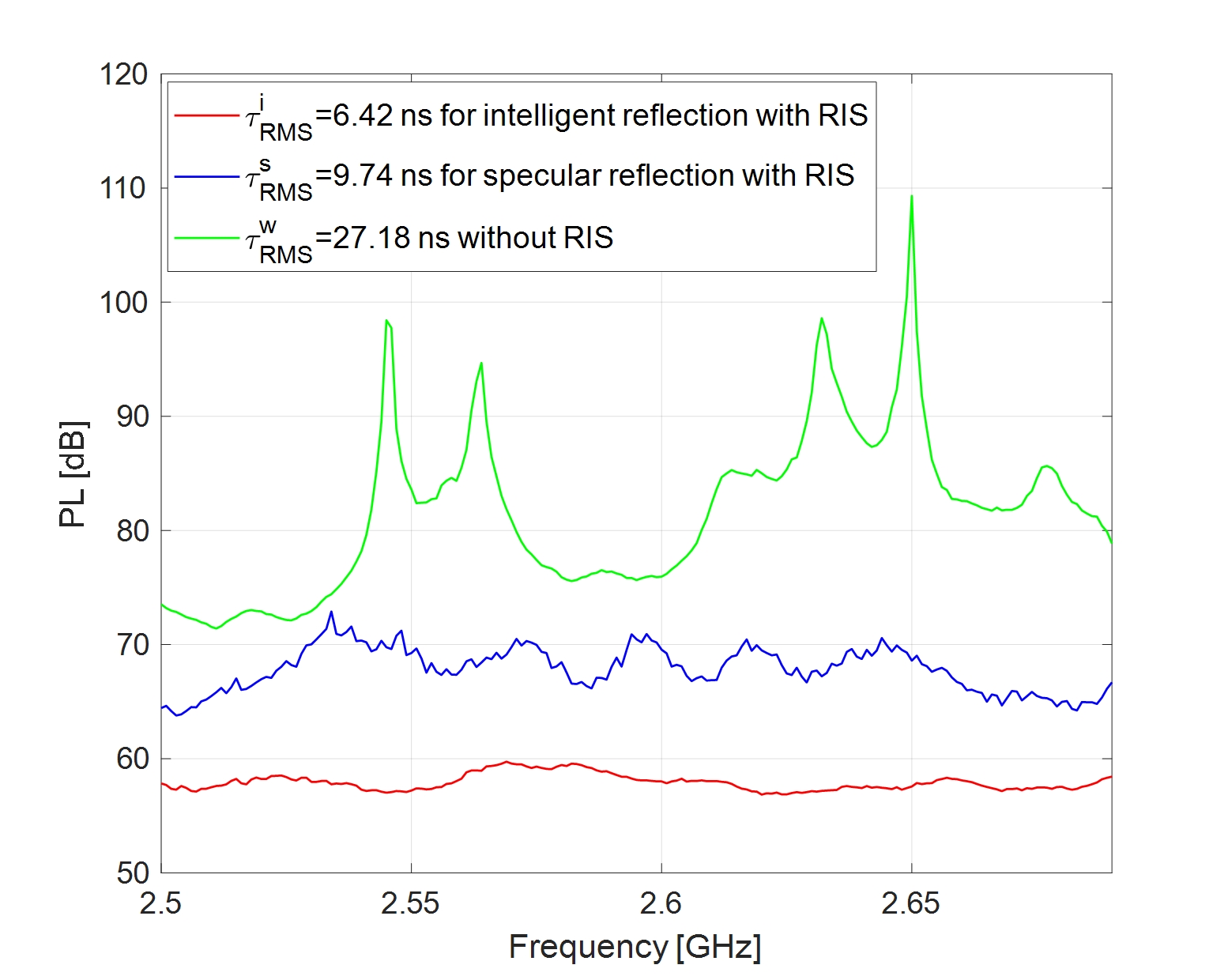}
\vspace{-0.5cm}
\caption{Frequency stationarity for three modes.}
\label{fig22}
\end{minipage}
\vspace{-0.5cm}
\end{figure}

\reffig{fig21} demonstrates the RMS DSs of three modes and the fitted Gaussian distributions $\mathcal{N}(\mu,\sigma^2)$, under $\theta_t = 45^\circ$ and $\theta_r = 45^\circ$. The means of these CDFs are $5.01$ ns, $5.88$ ns and $13.04$ ns for the modes of intelligent reflection, specular reflection and without RIS respectively, as displayed in \autoref{table_v1}. This phenomenon illustrates that the channel time dispersion of intelligent reflection is lower than that of specular reflection, because of the diffuse reflection of EM waves caused by the latter. The channel time dispersion without RIS is still the strongest,  on account of its dominant scattering, diffraction, and diffuse reflection. In addition, it should be noted that these three means in indoor corridor scenario are respectively lower than those ($8.42$ ns, $10.79$ ns and $37.55$ ns respectively) in outdoor scenario, which indicates that the signal energy is better focused when in corridor channels.

\reffig{fig22} demonstrates the frequency stationarity within the measured bandwidth for three modes under $\theta_t = 45^\circ$, $\theta_r = 45^\circ$, $d_1 = 8$ m and $d_2 = 8$ m. Similarly to \reffig{fig16}, the propagation signal without RIS shows the strongest frequency selective fading, followed by that of specular reflection with RIS. The intelligent reflection with RIS still has the flattest signal. Their RMS DSs are respectively $27.18$ ns, $9.74$ ns and $6.42$ ns, which accounts for the above phenomena theoretically according to the expression of \eqref{eq18}. This phenomenon indicates significantly the superiority of intelligent beamforming by RIS against ``plate metal'' or ``without RIS'', so as to support its claim of ``customizing wireless channels''.

\begin{table}[htbp]
\centering
\vspace{-0.2cm}
\begin{minipage}[t]{0.5\textwidth}
\caption{Fitting results for indoor measurement.}
\label{table6}
\vspace{-0.2cm}
\begin{tabular}{|c|c|c|c|}
\hline
\multirow{1}{*}{\tabincell{c}{Parameters}} & \multirow{1}{*}{\tabincell{c}{$\alpha$ ($PL_{FS}^{RIS}$)}} & \multirow{1}{*}{\tabincell{c}{$\beta_1$ ($n_1$)}} & \multirow{1}{*}{\tabincell{c}{$\beta_2$ ($n_2$)}} \\\hline
\multirow{2}{*}{\tabincell{c}{Modified FI with\vspace{-0.2cm}\\ \textbf{measurement} data}} & \multirow{2}{*}{\tabincell{c}{$28.16$}} & \multirow{2}{*}{\tabincell{c}{$1.9$}} & \multirow{2}{*}{\tabincell{c}{$1.68$}} \\
 &  &  &  \\\hline
\multirow{2}{*}{\tabincell{c}{Modified FI with\vspace{-0.2cm}\\ \textbf{free-space} data}} & \multirow{2}{*}{\tabincell{c}{$28.7$}} & \multirow{2}{*}{\tabincell{c}{$1.96$}} & \multirow{2}{*}{\tabincell{c}{$1.96$}} \\
 &  &  &  \\\hline
\multirow{2}{*}{\tabincell{c}{Modified CI with\vspace{-0.2cm}\\ \textbf{measurement} data}} & \multirow{2}{*}{\tabincell{c}{$26.97$}} & \multirow{2}{*}{\tabincell{c}{$1.95$}} & \multirow{2}{*}{\tabincell{c}{$1.74$}} \\
 &  &  &  \\\hline
\multirow{2}{*}{\tabincell{c}{Modified CI with\vspace{-0.2cm}\\ \textbf{free-space} data}} & \multirow{2}{*}{\tabincell{c}{$26.97$}} & \multirow{2}{*}{\tabincell{c}{$2.04$}} & \multirow{2}{*}{\tabincell{c}{$2.04$}} \\
 &  &  &  \\\hline
\end{tabular}
\end{minipage}
\begin{minipage}[t]{0.4\textwidth}
\caption{Fitting results for O2I measurement.}
\label{table7}
\vspace{-0.2cm}
\begin{tabular}{|c|c|c|} \hline
Parameters & $\alpha$ ($PL_{FS}^{RIS}$) & $\beta_2$ ($n_2$) \\ \hline
\multirow{2}{*}{\tabincell{c}{Modified FI with\vspace{-0.2cm}\\ \textbf{measurement} data}} & \multirow{2}{*}{\tabincell{c}{$46.11$}}  & \multirow{2}{*}{\tabincell{c}{$1.51$}}  \\
& & \\\hline
\multirow{2}{*}{\tabincell{c}{Modified FI with\vspace{-0.2cm}\\ \textbf{free-space} data}} & \multirow{2}{*}{\tabincell{c}{$47.64$}}  & \multirow{2}{*}{\tabincell{c}{$1.8$}}  \\
& & \\\hline
\multirow{2}{*}{\tabincell{c}{Modified CI with\vspace{-0.2cm}\\ \textbf{measurement} data}} & \multirow{2}{*}{\tabincell{c}{$46.06$}}  & \multirow{2}{*}{\tabincell{c}{$1.52$}}  \\
& & \\\hline
\multirow{2}{*}{\tabincell{c}{Modified CI with\vspace{-0.2cm}\\ \textbf{free-space} data}} & \multirow{2}{*}{\tabincell{c}{$46.06$}}  & \multirow{2}{*}{\tabincell{c}{$2$}}  \\
& & \\\hline
\end{tabular}
\end{minipage}
\vspace{-0.5cm}
\end{table}

\subsection{O2I measurement}

\begin{figure}[htbp]
\vspace{-0.5cm}
\centering
\begin{minipage}[t]{0.48\textwidth}
\centering
\includegraphics[width=3in]{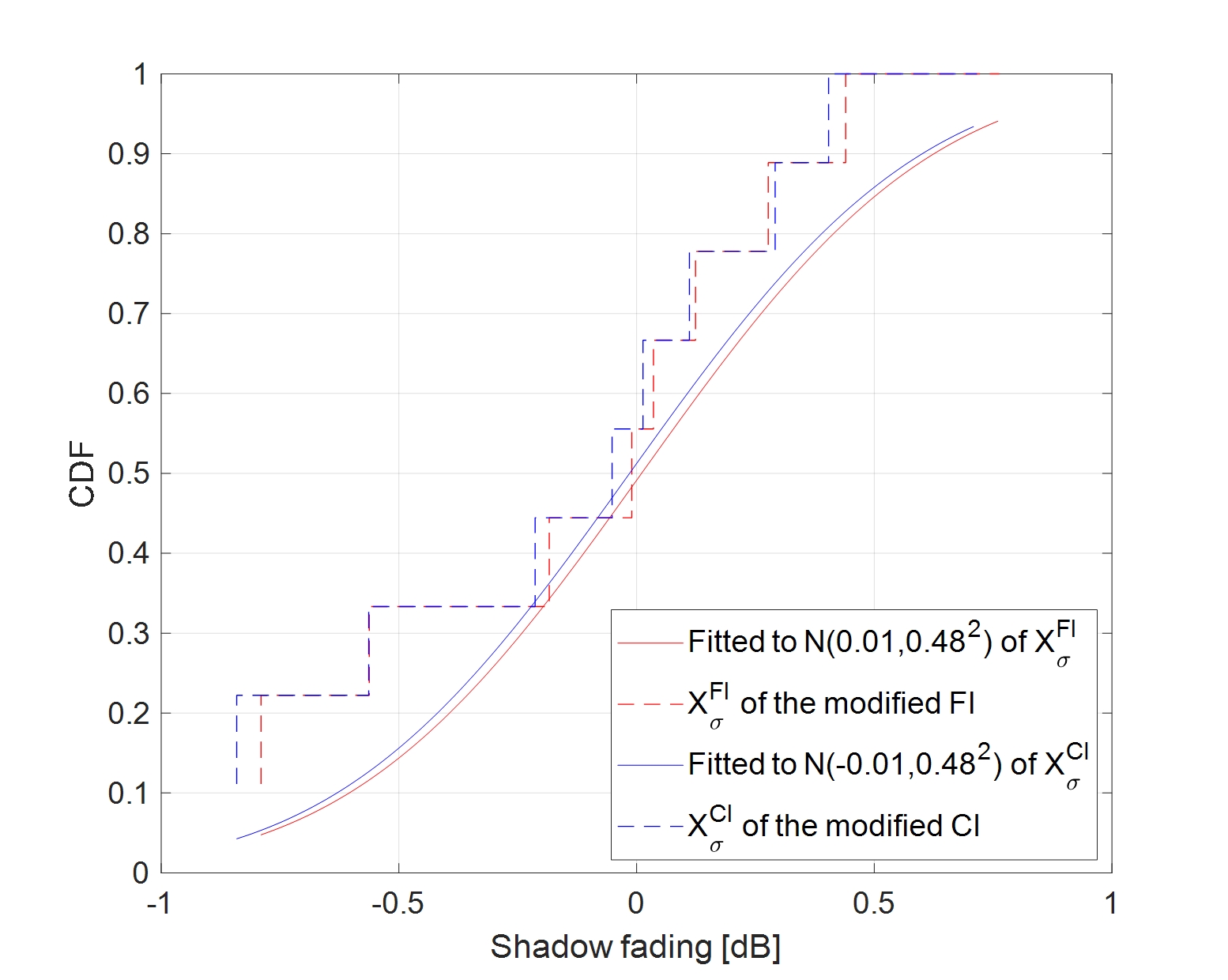}
\vspace{-0.5cm}
\caption{SFs and the fitted CDFs in left aisle.}
\label{fig23}
\end{minipage}
\begin{minipage}[t]{0.48\textwidth}
\centering
\includegraphics[width=3in]{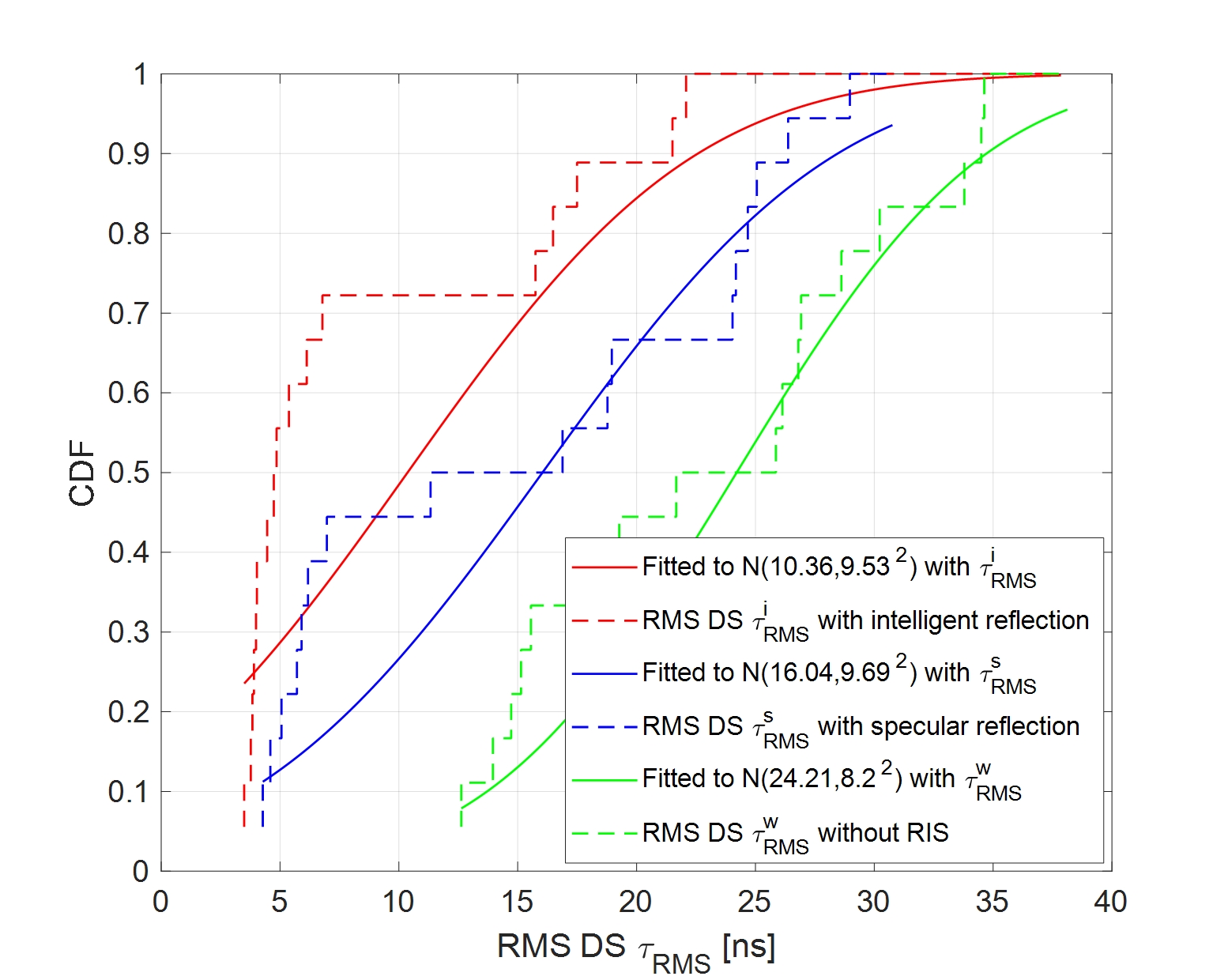}
\vspace{-0.5cm}
\caption{Comparison of RMS DSs for three modes.}
\label{fig27}
\end{minipage}
\vspace{-0.5cm}
\end{figure}

For the intelligent reflection mode in O2I measurement, the fitting results of the modified CI and FI models with the measured data as well as the free-space data are respectively summarized in \autoref{table7}. Note that only the measured data in left aisle is fitted to the models, where $\theta_t=45^\circ$, $\theta_r=45^\circ$, and $d_1=9$ m are fixed and the PLE on $d_2$ is taken into account in \autoref{table7}. The reference value $PL_{FS}^{RIS} = 46.06$ of the modified CI model is calculated under $\theta_t = 45^\circ$, $\theta_r = 45^\circ$, $d_1 = 9$ m, and $d_2 = 1 $ m. From this table, we find that for both of the modified CI and FI models, the PLEs on $d_2$ with free-space data are significantly higher than those with measurement data. This results from that in O2I measurement, there exists plentiful reflection and scattering caused by the surroundings such as desks and chairs. \reffig{fig23} exhibits the CDFs of SFs for the modified models, both of which are fitted into the Gaussian distributions with $\sigma = 0.48$, indicating their similar and high prediction accuracy.

\reffig{fig27} shows the RMS DSs of three modes and the fitted Gaussian distributions $\mathcal{N}(\mu,\sigma^2)$. Their means are $10.36$ ns, $16.04$ ns and $24.21$ ns respectively for intelligent reflection, specular reflection and without RIS, as shown in \autoref{table_v1}. The channel time dispersion of intelligent reflection is still the lowest due to its property of focusing energy. Note that these three means in classroom scenario are higher than those ($8.42$ ns, $10.79$ ns and $37.55$ ns respectively) in outdoor scenario  and those ($5.01$ ns, $5.88$ ns and $13.04$ ns respectively) in corridor scenario. This indicates that the time dispersion for RIS-assisted channel in O2I scenario is strongest.

\begin{figure*}[htbp]
\vspace{-0.5cm}
     \centering
     \subfloat[]{\includegraphics[height = 2.5in]{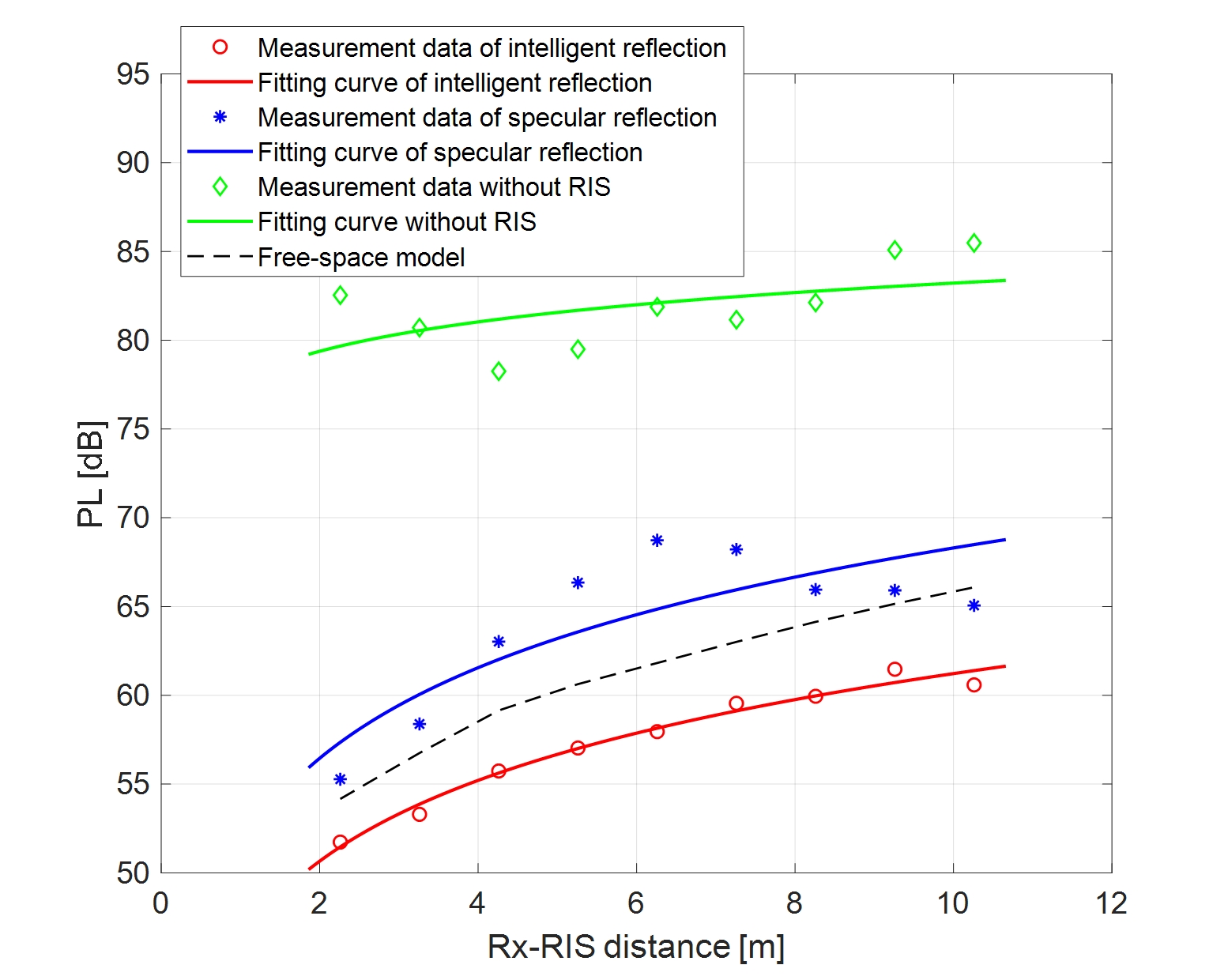} \hfill \label{fig24a}}
     \subfloat[]{\includegraphics[height = 2.5in]{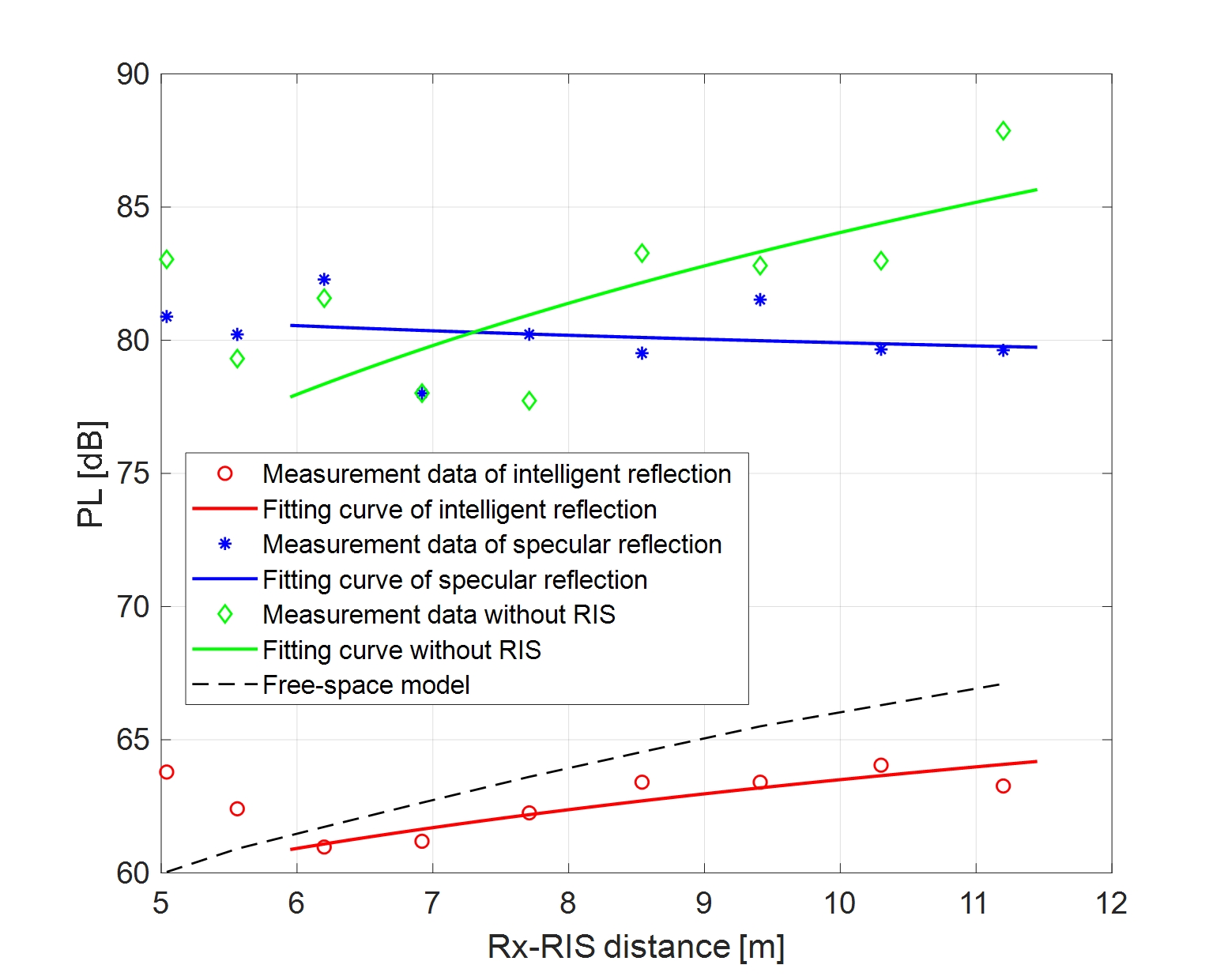} \hfill \label{fig24b}}
     \vspace{-0.2cm}
    \caption{PL comparison of measurement data and free-space data. (a) In left aisle. (b) In right aisle. }
\label{fig24}
\vspace{-0.5cm}
\end{figure*}

\reffig{fig24} shows the measured PLs for three modes in left and right aisles respectively as well as the free-space PLs. Note that in right aisle, the Positions 1 and 2 are discarded because their propagation links with RIS are blocked by the teacher’s table during the measurement. In left aisle shown in \reffig{fig24a}, the PLE on $d_2$ of intelligent reflection mode is slightly lower than that of free-space PLE, resulting from the abundant scatters in this classroom. In addition, the PL for intelligent reflection demonstrates an improvement of about $7$ dB than that for specular reflection and an improvement of about $25\sim30$ dB than that without RIS. From \reffig{fig24b}, the PLE on $d_2$ of intelligent reflection mode is greatly lower than that of free-space PLE in right aisle. The PL for intelligent reflection mode demonstrates an improvement of about $15\sim20$ dB than that with specular reflection mode and an improvement of about $20\sim25$ dB than that without RIS. 

\begin{figure*}[htbp]
\vspace{-1cm}
     \centering
     \subfloat[]{\includegraphics[height = 2.5in]{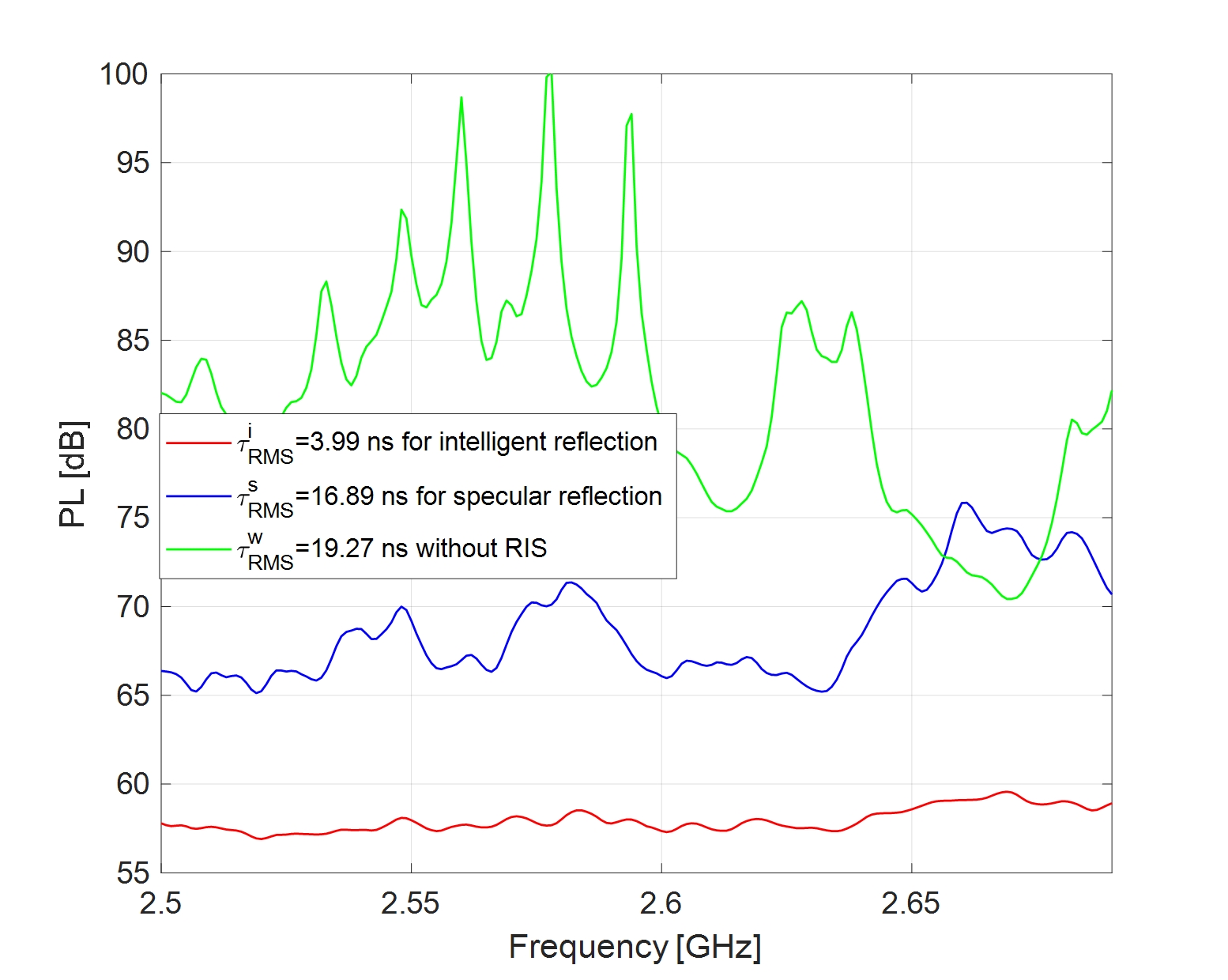}\hfill \label{fig28a}}
     \subfloat[]{\includegraphics[height = 2.5in]{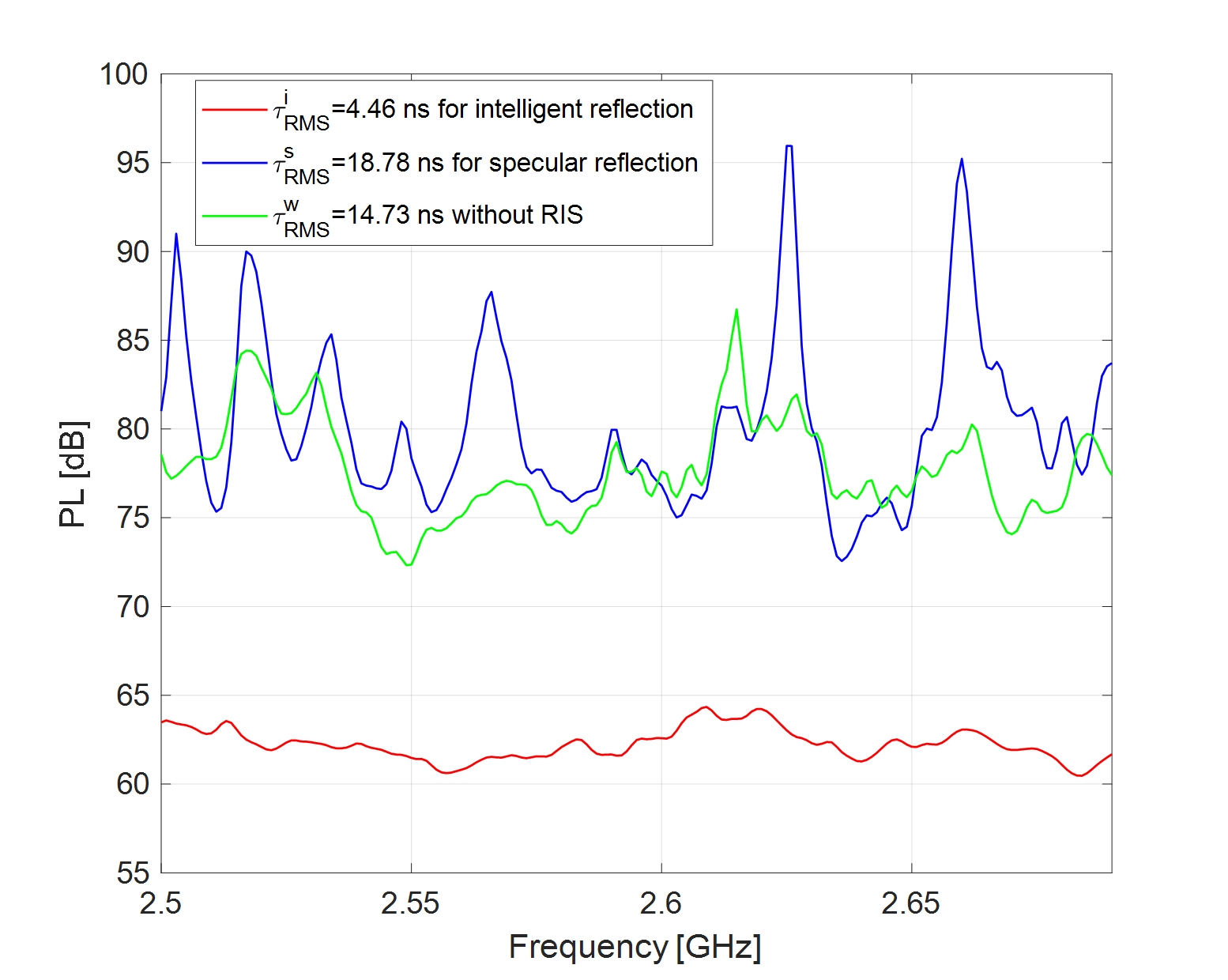}\hfill \label{fig28b}}
     \vspace{-0.2cm}
    \caption{Frequency stationarity for three modes. (a) In left aisle. (b) In right aisle.}
\label{fig28}
\vspace{-0.5cm}
\end{figure*}

\reffig{fig28} demonstrates the frequency stationarity within the measured bandwidth for three modes at Position 5 in left and right aisles, respectively. From \reffig{fig28a}, in left aisle, the RMS DSs are $\tau _{\rm{RMS}}^{\rm{i}}=3.99$ ns, $\tau_{\rm{RMS}}^{\rm{s}}=16.89$ ns and $\tau_{\rm{RMS}}^{\rm{w}}=19.27$ ns respectively for the modes of intelligent reflection, specular reflection as well as without RIS. Their signals are respectively flat, slight frequency selective fading, and strong frequency selective fading. As shown in \reffig{fig28b}, in right aisle, the frequency stationarity for the modes of intelligent reflection and without RIS are similar to those in \reffig{fig28a}. Nevertheless, the specular reflection mode in this figure illustrates a stronger frequency selective fading than that in left aisle. This  attributes to that the specular reflection mode is incapable of covering the Rx positions in right aisle directly, where only the reflected and scattered signals arrive at Rx.

\vspace{-0.2cm}
\subsection{Summary of measurement and modeling}

\noindent{\em 1) Channel measurement}

On \textbf{PL}: In indoor and outdoor measurement, the PLEs on both distance and angle for intelligent reflection mode are approximate to the free-space PLEs. In O2I measurement, the PLE with the measured data is prominently lower than that in free space, which may result from the abundant scattering in classroom. In addition, a slight gap between the measured PL and free-space PL is observed. In all of the three measurements, the intelligent reflection with RIS shows a significant improvement on PL covering multiple desired directions, against that of specular reflection mode as well as the mode without RIS.

On \textbf{RMS DS}: For intelligent reflection with RIS, specular reflection with RIS, and the mode without RIS, their means of RMS DSs are respectively $10.36$ ns, $16.04$ ns and $24.21$ ns in O2I scenario, $8.42$ ns, $10.79$ ns and $37.55$ ns in outdoor scenario, as well as $5.01$ ns, $5.88$ ns and $13.04$ ns in indoor scenario. This indicates that for RIS-assisted channel, the time dispersion is the strongest in O2I scenario, followed by outdoor scenario, and the weakest time dispersion occurs in indoor channel. Moreover, the intelligent reflection is evidently favorable for focusing signal energy and reducing time dispersion. 

On \textbf{frequency stationarity}: In all of the three measurements, the intelligent reflection with RIS demonstrates a flat propagation signal. By contrast, the specular reflection with RIS shows a slight but non-negligible frequency selective fading and the propagation mode without RIS manifests the strongest frequency selective fading. These phenomena evidence that the intelligent reflection with RIS is advantageous for enhancing frequency stationarity and may pave the way for promoting RIS-empowered high data rate transmission. 

On \textbf{spatial consistency}: The MPC evolution for intelligent reflection mode shows a stationary spatial consistency as the variation of distance. In addition, plenty of additional MPCs are observed for specular reflection mode. For the mode without RIS, a nonstationary spatial consistency with weak power is detected. This may have implications on the future research such as RIS-empowered beam tracking. \\

\vspace{-0.2cm}
\noindent{\em 2) Channel modeling}

On \textbf{modeling accuracy}: Both of the the modified FI and CI models can be viewed as accurate and generalizable models to predict the PL characteristics of RIS-assisted channel covering distance and angle domains. Their fitting standard deviations are $2.53$ dB in outdoor scenario, $2.46$ dB in indoor scenario, as well as $0.48$ dB in O2I scenario. This phenomenon illustrates that for RIS-assisted channel, the proposed two models show a similar and high prediction accuracy. These results may has the instructive significance for future RIS-related PL evolution as well as modeling standardization.

On \textbf{modeling complexity}: The proposed two empirical models are inherited from the traditional FI and CI models respectively, which have been widely considered in reports, specifications and studies. The modified FI and CI models only parameterize several variables with the formal expression unchanged, while more comprehensively predicting the propagation characteristics for RIS-assisted channel. Thus, the modeling complexity can be reasonably acceptable and widely applicable for RIS-related practical deployment in the future.

\section{Conclusions}
\label{sec6}
In this paper, we conducted multi-scenario broadband channel measurement campaigns and channel modeling for RIS-assisted SISO channel at 2.6 GHz. Utilizing the VNA-based measurement system and a fabricated RIS, three measurements including outdoor, indoor and O2I scenarios were carried out, with 2096 channel acquisitions collected. In each scenario, three propagation modes including intelligent reflection with RIS, specular reflection with RIS, and the mode without RIS were measured. In addition, two modified empirical models including FI and CI, were proposed to cater for the cascaded characteristics of RIS-assisted channel. From the measurement results, the PL for ``intelligent reflection with RIS'' is significantly lower than that for ``specular reflection with RIS'' and that for ``without RIS'', which manifests its potential of improving the channel quality. The PLEs in outdoor and indoor scenarios show similar values as those in free space. Nevertheless, the PLE in O2I scenario is prominently lower than the free space value, which may be due to the plentiful scattering and reflection of the surroundings. Furthermore, for RIS-assisted channel, RMS DSs are highest in O2I scenario, followed by outdoor scenario, and the RMS DSs in indoor scenario are lowest. From the perspective of propagation mode, the intelligent reflection is more favorable for focusing signal energy and reducing time dispersion. Additionally, the measurement results also illustrate that the intelligent reflection with RIS is beneficial to enhancing frequency stationarity. The MPC evolution for intelligent reflection mode shows a more stationary spatial consistency. In our future studies, more diverse yet typical RIS-assisted communication scenarios such as urban macro cell (UMa) and MIMO, will be further measured and verified.



\vfill

\end{document}